\documentclass[12pt, letterpaper]{article}
\usepackage[margin=1in]{geometry}
\usepackage[usenames,dvipsnames,svgnames]{xcolor}
\definecolor{red}{rgb}{0.8,0.0,0.0}
\usepackage{amsmath, amssymb, siunitx, slashed, esvect, tikz, cite, rotating, setspace, IEEEtrantools, graphicx, mathtools, caption, tikz-feynman, subcaption}
\usepackage[colorlinks=true,pdfborder={0 0 0},citecolor=blue,linkcolor=blue]{hyperref}
\usetikzlibrary{quantikz2}
\newcommand{\volkovprop}{double, double distance=2pt, with arrow=0.5, arrow size=2pt}

\graphicspath{ {./Images/} }

\definecolor{green}{rgb}{0.0,0.4,0.1}
\definecolor{darkblue}{rgb}{0.0,0.1,0.7}

\renewcommand{\bar}[1]{\overline{#1}}
\renewcommand{\vec}[1]{\vv{#1}}
\renewcommand{\tilde}[1]{\widetilde{#1}}
\renewcommand{\ket}[1]{\lvert #1 \rangle}
\renewcommand{\bra}[1]{\langle #1 \rvert}

\newcommand{\expect}[1]{\langle #1 \rangle}
\DeclareMathOperator{\diag}{diag}

\title{Quantum Simulations for Strong-Field QED}
\author{Luis Hidalgo \and Patrick Draper \and \small{\textit{Department of Physics, University of Illinois Urbana-Champaign}} \and \small{\textit{Illinois Quantum Information Science and Technology Center}}}
\date{}

\begin{document}
	
	\maketitle
	
	\begin{abstract}
		Quantum field theory in the presence of strong background fields contains interesting problems where  quantum computers may someday provide a valuable computational resource. In the NISQ era it is useful to consider simpler benchmark problems in order to develop feasible approaches, identify critical limitations of current hardware, and build new simulation tools. Here we perform quantum simulations of strong-field QED (SFQED) in 3+1 dimensions, using real-time nonlinear Breit-Wheeler pair-production as a prototypical process. The strong-field QED Hamiltonian is derived and truncated in the Furry-Volkov mode expansion, and the interactions relevant for Breit-Wheeler are transformed into a quantum circuit. Quantum simulations of a ``null double slit" experiment are found to agree well with classical simulations following the application of various error mitigation strategies, including an asymmetric depolarization algorithm  which we develop and adapt to the case of Trotterization with a time-dependent Hamiltonian. We also discuss longer-term goals for the quantum simulation of SFQED.
		
	\end{abstract}
	
	\section{Introduction}
	
	Quantum computing technology is rapidly developing. At present, real quantum devices do not offer useful  advantage over classical computers, but in the not terribly distant future, quantum computing may mature into a valuable tool with diverse applications. In high energy physics (HEP), it is hoped that quantum computers will be able to simulate phenomena in models where classical simulation techniques are limited by sign problems or the need to explore a large Hilbert space \cite{Klco_2022, humble2022snowmass, bauer2023}. In the near term, it is important to develop the tools to maximize the power of NISQ devices on simulation problems, and to flesh out the space of target problems for which systematic improvements in quantum computing power would lead to clear advantage over classical simulations \cite{PRXQuantum.4.027001, dimeglio2023quantum}.
	
	The Schwinger process, $e^+e^-$ pair creation in 1+1 dimensions in a strong background electric field \cite{PhysRev.128.2425}, is a widely used benchmark for quantum simulations in HEP (see, e.g., \cite{Shaw2020quantumalgorithms, PhysRevX.3.041018, PhysRevA.98.032331, Martinez_2016, Kokail_2019, PRXQuantum.3.020324, PRXQuantum.4.030323, de_Jong_2022, Pomarico_2023, Surace_2020, Nagano_2023}.) As a nonperturbative, dynamical process in a confining gauge theory, it captures some of the essential physics of  four-dimensional QCD-like theories in a simpler setting more tractable for NISQ-era devices. The potential applications of quantum computing to QCD are well known \cite{PhysRevResearch.5.033184, PhysRevD.107.054512, PhysRevD.107.054513, ciavarella2023quantum, PhysRevD.103.094501, Ciavarella_2022}. Perhaps less widely appreciated are the potential applications of quantum simulation to QED, particularly in the presence of strong background electromagnetic fields (commonly referred to as strong-field quantum electrodynamics, or SFQED -- for a recent review, see \cite{FEDOTOV20231}.) Indeed the Schwinger process \cite{PhysRev.82.664, PhysRevLett.96.140402, PhysRevD.78.061701} is a simple example of an SFQED phenomenon, but the richness and complexity of the theory arises in the presence of dynamical photons. Remarkably, although QED near the perturbative vacuum has been tested with unparalleled precision, there are still aspects of QED -- most dramatically in the regime of ultra-strong fields -- where the behavior is not fully  understood (see e.g. \cite{Fedotov_2017,FEDOTOV20231}). Qualitatively, ultra-strong fields should yield explosive particle production and backreaction, leading rapidly to a complicated quantum state. Thus quantum computers may one day provide a unique probe of the most extreme regimes of QED. Similarly, quantum computers could provide a valuable tool to study phenomena in moderately intense fields. In this regime, which will be probed in upcoming experiments \cite{jacobs2022luxe}, analytic techniques are available for studying few-particle scattering amplitudes, at low loop order, in idealized background fields. As in QCD, quantum simulations could provide complementary access to real-time phenomena in more complex states. 
	
	These are future goals. Near-term quantum computers will be limited by size, connectivity, and noise. Therefore the present task is to design general simulation frameworks and test them on simple benchmark processes, tractable on present-day hardware. 
	
	In this paper we study real-time digital quantum simulations of 3+1D SFQED. Unlike the discretized-space approach generally used to study Hamiltonian lattice gauge theories \cite{PhysRevA.73.022328, PhysRevD.11.395}, we use the discretized momentum-space SFQED Hamiltonian in the light-front formalism \cite{Valdés_2004, Zhao_2013, PhysRevD.1.2901}.  We  perform quantum simulations by mapping Fock states onto qubits and ladder operators onto Pauli gates. As a simple but nontrivial benchmark process, we examine nonlinear Breit-Wheeler pair-production \cite{PhysRev.46.1087, PhysRevLett.108.240406, PhysRevA.86.052104, Blackburn_2018, PhysRevA.105.013105, PhysRevD.91.013009, Titov_2020, PhysRevD.94.013010}. In this process, a photon ``decays'' into an electron-positron pair via an interaction with a strong background electromagnetic field. We study the real-time interference effects in pair-production probability due to two fields separated in time, a sort of ``null double-slit" experiment~\cite{Ilderton_2020, Ilderton_2019, Granz_2019, Jansen_2017, PhysRevA.90.052108}.
	
	Currently, quantum noise is unavoidable in simulations, and error mitigation is a powerful tool to ``fit and subtract" various sources of noise. Fortunately, the last decade has seen rapid progress in the development of a suite of effective mitigation techniques \cite{Nation_2021, Souza_2012, PhysRevA.94.052325, cai2023quantum, Temme_2017, Giurgica_Tiron_2020}. Here we employ measurement mitigation, Pauli twirling \cite{PhysRevA.94.052325}, and depolarization mitigation. For depolarization mitigation, we improve on so-called ``self-mitigation'' \cite{rahman2022real, PhysRevLett.127.270502, PhysRevD.107.054512} by relaxing the symmetric depolarization assumption. We find that at present these techniques are essential to obtain acceptable results. Without the combination of these mitigation techniques, our quantum simulations would yield unusable data.
	
	This paper is organized as follows. In Sec. \ref{sec_sfqed_hamiltonian} we derive the momentum-space SFQED Hamiltonian and describe the truncation approach we use for quantum simulations. In Sec. \ref{sec_quantum_circuit_design} we translate the nonlinear Breit-Wheeler process into quantum circuits. In Sec. \ref{sec_breit_wheeler} we describe the specific benchmark process of interest and develop a  theoretical understanding of the relevant interference phenomena using time-dependent perturbation theory. We also perform real-time classical simulations in order to compare with quantum simulation results. In Sec. \ref{sec_quantum_results} we show raw quantum data and its comparison with Qiskit noisy simulations. In Sec. \ref{sec_error_mitigation} we describe the error mitigation strategies used to obtain the final mitigated results, which are found to be in good agreement with the classical simulations. Finally, in Sec. \ref{sec_conclusion} we describe some directions for future work.
	
	\subsection{Conventions}
	
	We work in natural units with $\hbar=c=1$ and  report values in $\si{\mega\electronvolt}$. $m=\SI{0.511}{\mega\electronvolt}$ is the mass of the electron and $e=0.303$ is the electric charge.
	
	It is convenient to work in light-front coordinates \cite{Zhao_2013, harindranath1998, RevModPhys.21.392}. The coordinate four-vector is $x^\mu = (x^+,x^-,x^1,x^2) = (x^+,x^-,x^\perp)$, where $x^\pm \equiv x^0 \pm x^3$. Similar notation holds for other four-vectors. The Minkowski metric and inverse are 
	\begin{equation}
		g_{\mu\nu} = \begin{pmatrix}
			0 & \frac{1}{2} & 0 & 0 \\
			\frac{1}{2} & 0 & 0 & 0 \\
			0 & 0 & -1 & 0 \\
			0 & 0 & 0 & -1
		\end{pmatrix}\qquad g^{\mu\nu} = \begin{pmatrix}
			0 & 2 & 0 & 0 \\
			2 & 0 & 0 & 0 \\
			0 & 0 & -1 & 0 \\
			0 & 0 & 0 & -1
		\end{pmatrix}.
		\label{metric}
	\end{equation}
	To alleviate the subsequent bookkeeping of factors of two or one-half, we work mainly with raised indices, so that $x_\mp \to \frac{x^\pm}{2}$ and $x_i \to -x^i$ (for $i=1,2$). 
	In other words, the covariant coordinate form is $x_\mu = (x_+,x_-,x_\perp) = (\frac{x^-}{2}, \frac{x^+}{2}, -x^\perp)$. An exception to this convention is with the four-gradient, which will be written as $\partial_\mu = (\partial_+, \partial_-, \partial_\perp)$.
	
	We take $x^+$ to be the light-front time coordinate. The spatial 3-vector and kinetic momentum are
	\begin{equation}
		\mathsf{x} = (x^-, x^1, x^2) \qquad \mathsf{p} = (p^+, p^1, p^2).
	\end{equation}
	We define their inner product as $\mathsf{p}\mathsf{x} = \frac{1}{2}p^+x^- - p^1x^1 - p^2x^2$. $p^-$ is light-front energy, and free particles satisfy the dispersion relation
	\begin{equation}
		p^- = \frac{(p^\perp)^2 + m^2}{p^+}.
	\end{equation}
	In the light-front, $p^+$ and hence $p^-$ are positive semidefinite. We will work in a Fock space constructed from a momentum lattice, and the zero mode $p^+=0$ does not propagate. Therefore we will take $p^+>0$. 
	
	Due to the conventions in Eq.~\eqref{metric}, the  integration measure on a slice of constant light-front time $x^+$ incurs a factor of one-half:
	\begin{equation}
		d^3\mathsf{x} = \frac{1}{2} dx^- dx^\perp.
	\end{equation}
	We work with the chiral basis for the gamma matrices, so that with $\mu=0,1,2,3$
	\begin{equation}
		\gamma^\mu = \begin{pmatrix}
			0 & \sigma^\mu \\
			\bar{\sigma}^\mu & 0
		\end{pmatrix}
	\end{equation}
	where $\sigma^\mu = (I,\vec{\sigma})$ and $\bar{\sigma}^\mu = (I,-\vec{\sigma})$. Light-front gamma matrices are defined as $\gamma^\pm = \gamma^0 \pm \gamma^3$. They obey the usual anticommutation relations with the light-front metric, $\{ \gamma^\mu, \gamma^\nu \} = 2g^{\mu\nu}$.
	
	\section{SFQED Hamiltonian}
	\label{sec_sfqed_hamiltonian}
	
	Given a quantum state $\ket{\psi}$, the generator of light-front time $x^+$ translations is $P_+$, $i\partial_+\ket{\psi} = P_+\ket{\psi}$. Equivalently, $\partial_+\ket{\psi} = -\frac{i}{2}P^-\ket{\psi}$ where we identify $P^- \equiv H$ as the Hamiltonian. The Hamiltonian density is given by the Legendre transform
	\begin{equation}
		\mathcal{H} = \frac{\partial\mathcal{L}}{\partial(\partial_+ \varphi_\alpha)} \partial_+ \varphi_\alpha - \mathcal{L},
		\label{legendre}
	\end{equation}
	where $\mathcal{L}$ is the  Lagrangian density
	\begin{equation}
		\mathcal{L} = -\frac{1}{4}F_{\mu\nu}F^{\mu\nu} + \bar{\psi}(i\slashed{D} - m)\psi.
		\label{lagrangian}
	\end{equation}
	Within the covariant derivative $D_\mu = \partial_\mu + ieA_\mu + ie\mathcal{A}_\mu$ we make explicit the electromagnetic gauge field $A_\mu$ and the classical\footnote{Eq. (\ref{lagrangian}) lacks a kinetic term for the background field $\mathcal{A}_\mu$ because we consider cases where it obeys Maxwell's equations in vacuum.} 
	background field $\mathcal{A}_\mu$. $\psi$ is the Dirac field and the Dirac adjoint is defined as usual by $\bar{\psi} = \psi^\dag \gamma^0$.
	
	We work in the gauge $A^+=0$ for the photon field and assume the Lorenz gauge $\partial_\mu \mathcal{A}^\mu = 0$ for the background field. For plane waves, $\mathcal{A}_\mu$ is a function of a light-front wavevector $\kappa^\mu$: $\mathcal{A}_\mu = \mathcal{A}_\mu(\kappa_\mu x^\mu)$.  If the background field is a null field  propagating in the $-\hat{z}$ direction,  $\kappa^\mu = (0,2\omega,0,0)$, then the Lorenz gauge reduces to $\mathcal{A}^+=0$. Here $\omega$ is the frequency of the background wave.
	
	Under these gauge conditions the Euler-Lagrange equations for $A_\mu$ and $\psi$ reveal two constraint equations. For the electromagnetic field, we have
	\begin{equation}
		\partial_-^2 A^- = -\partial_- \partial_i A^i -\frac{1}{2} e\bar{\psi}\gamma^+\psi.
		\label{photon_constraint}
	\end{equation}
	For the Dirac field, we obtain the Dirac equation $(i\slashed{D}-m)\psi=0$, but defining the projectors
	\begin{equation}
		\Lambda_+ = \frac{1}{4}\gamma^+ \gamma^- = \diag(1,0,0,1) \qquad \Lambda_- = \frac{1}{4}\gamma^- \gamma^+ = \diag(0,1,1,0),
		\label{projectors}
	\end{equation}
	and applying $\Lambda_-$ from the left, we obtain the constraint
	\begin{equation}
		\partial_-\psi_+ = \frac{i}{4}\gamma^+(i\gamma^i\partial_i + e\gamma^i A^i + e\gamma^i \mathcal{A}^i - m)\psi_-
		\label{fermi_constraint}
	\end{equation}
	where $\psi_\pm = \Lambda_\pm \psi$ and $\psi_+ + \psi_- = \psi$. We will solve the constraints exactly for $\psi_+$ and $A^-$,
	\begin{equation}
		\psi_+ = \frac{i \gamma^+(i\gamma^i\partial_i + e\gamma^i A^i + e\gamma^i \mathcal{A}^i - m)\psi_-}{4 \partial_-} \qquad A^- = -\frac{\partial_- \partial_i A^i + e\psi_-^\dag \psi_-}{\partial_-^2}.
		\label{constraints}
	\end{equation}
	Note that $\gamma^0 \gamma^+ = 2\Lambda_-$ and $\Lambda_-^2 = \Lambda_-$, resulting in the appearance of $\psi_-$ in the solution for $A^-$.
	
	In defining the Green functions appearing in Eq.~(\ref{constraints}), it is customary to take antisymmetric boundary conditions for fields at the longitudinal boundaries~\cite{PhysRevD.1.2901, Valdés_2004, heinzl_1994}. As a consequence the zero mode is omitted from the spectral decomposition. 
	
	Having fixed the gauge and solved the constraints exactly, the Hilbert space we construct is a physical one, with no additional non-gauge-invariant sectors. This approach minimizes the number of qubits needed to represent the degrees of freedom, at the expense of introducing additional interactions.
	
	Inserting Eqs.~\eqref{constraints} into Eq.~\eqref{legendre} we obtain a nonlocal expression for the Hamiltonian density in terms of the dynamical fields $A^i$ and $\psi_-$. The full Hamiltonian density can be separated into several terms: the fermion energy
	\begin{equation}
		\mathcal{H}_\psi = -\frac{im^2}{2}\psi_-^\dag\left( \frac{\psi_-}{\partial_-} \right) + \frac{i}{2}\psi_-^\dag\left( \frac{\partial_i\partial_i\psi_-}{\partial_-} \right),
	\end{equation}
	the photon energy
	\begin{equation}
		\mathcal{H}_A = \frac{1}{2}(\partial_1 A^2 - \partial_2 A^1)^2 - \frac{1}{2}(\partial_-\partial_i A^i)\left( \frac{\partial_i A^i}{\partial_-} \right),
	\end{equation}
	the fermion-background energy
	\begin{equation}
		\mathcal{H}_{\psi\mathcal{A}} = e\mathcal{A}^- \psi_-^\dag \psi_- + e \mathcal{A}^i \psi_-^\dag \left( \frac{\partial_i \psi_-}{\partial_-} \right) - \frac{ie^2}{2} \psi_-^\dag \mathcal{A}^i \mathcal{A}^i \left( \frac{\psi_-}{\partial_-} \right),
	\end{equation}
	the fermion-photon-background interation
	\begin{equation}
		\mathcal{H}_{\psi A\mathcal{A}} = \frac{ie^2}{2} \psi_-^\dag A^i \gamma^i \gamma^j \mathcal{A}^j \left( \frac{\psi_-}{\partial_-} \right) + \frac{ie^2}{2} \psi_-^\dag \mathcal{A}^i \gamma^i \gamma^j \left( \frac{A^j \psi_-}{\partial_-} \right),
	\end{equation}
	the fermion-photon interaction
	\begin{IEEEeqnarray*}{rCl}
		\mathcal{H}_{\psi A} & = & \frac{ime}{2} \psi_-^\dag \gamma^i \left( \frac{A^i\psi_-}{\partial_-} \right) - \frac{ime}{2} \psi_-^\dag \gamma^i A^i \left( \frac{\psi_-}{\partial_-} \right) \\
		& - & \frac{e}{2} (\partial_- \partial_i A^i)\left( \frac{\psi_-^\dag \psi_-}{\partial_-^2} \right) - \frac{e}{2} (\psi_-^\dag \psi_-) \left( \frac{\partial_i A^i}{\partial_-} \right) \\
		& - & \frac{e}{2} \psi_-^\dag \gamma^i \gamma^j \left( \frac{\partial_i(A^j \psi_-)}{\partial_-} \right) - \frac{e}{2} \psi_-^\dag A^i \gamma^i \gamma^j \left( \frac{\partial_j \psi_-}{\partial_-} \right),
	\end{IEEEeqnarray*}
	the four-fermion interaction
	\begin{equation}
		\mathcal{H}_{4\psi} = - \frac{e^2}{2} (\psi_-^\dag \psi_-) \left( \frac{\psi_-^\dag \psi_-}{\partial_-^2} \right),
	\end{equation}
	and the double fermion-photon interaction
	\begin{equation}
		\mathcal{H}_{2\psi A} = \frac{ie^2}{2} \psi_-^\dag A^i \gamma^i \gamma^j \left( \frac{A^j \psi_-}{\partial_-} \right).
	\end{equation}
	
	\subsection{Momentum-Space Hamiltonian and Discretization}
	Most simulations of quantum field theories on quantum computers use a real-space lattice discretization. We instead use a mode decomposition and construct the corresponding Fock space in the usual way. This approach has the advantage that it makes extracting physical quantities much easier with noisy, low-qubit-number devices. In terms of asymptotic scaling, the mode expansion results in a Hamiltonian with $O(n^4)$ terms, compared with $O(n^3)$ terms for a spatial lattice.
	
	We  define creation and annihilation operators $a^\dag/a$, $b^\dag/b$, and $c^\dag/c$ for the electron, positron, and photon, respectively. These respect the  (anti)commutation relations
	\begin{equation}
		\{ a^s_p, a^{r\dag}_{p'} \} = (2\pi)^3 \delta_{s,r} \delta^3(\mathsf{p} - \mathsf{p}') \qquad \{ b^s_p, b^{r\dag}_{p'} \} = (2\pi)^3 \delta_{s,r} \delta^3(\mathsf{p} - \mathsf{p}') \qquad [ c^j_k, c^{j'\dag}_{k'} ] = (2\pi)^3 \delta_{j,j'} \delta^3(\mathsf{k} - \mathsf{k}').
		\label{commutators}
	\end{equation}
	Here, $s,r=\pm\frac{1}{2}$ are helicity indices for the fermions and $\lambda=1,2$ are polarization indices for the photon. The Fock states are labeled by occupation numbers and require an ordering, particularly for the fermions. We write the states in the following order:
	\begin{equation}
		\ket{\dots, {e^-}_{p_n}^{+\frac{1}{2}}, \dots, {e^+}_{p_n}^{+\frac{1}{2}}, \dots, \gamma_{p_n}^1, \dots, {e^-}_{p_n}^{-\frac{1}{2}}, \dots, {e^+}_{p_n}^{-\frac{1}{2}}, \dots, \gamma_{p_n}^2,\dots},
		\label{fock}
	\end{equation}
	where the superscript is the helicity/polarization, the subscript is the four-momentum, and $n$ is an index for all allowed four-momenta for a given particle and helicity/polarization. (For a given three-momentum, $p^-$ is determined by the dispersion relation.) We note that the fermion creation and annihilation operators induce a phase factor of $(-1)^\zeta$, where $\zeta$ is the sum of the number of fermions to the left of the operand as written in Eq. (\ref{fock}). The fermion states can be at most singly occupied, while the  photon states can be arbitrarily highly occupied.
	
	The mode expansion of the fields also requires helicity bispinors for the Dirac field and polarization vectors for the electromagnetic field. Since $\psi_-$ is projected with $\Lambda_-$, it is convenient to define basis bispinors for each helicity as
	\begin{equation}
		w^{+\frac{1}{2}} = \begin{pmatrix}
			0 \\ 1 \\ 0 \\ 0
		\end{pmatrix} \qquad w^{-\frac{1}{2}} = \begin{pmatrix}
			0 \\ 0 \\ 1 \\ 0
		\end{pmatrix}.
	\end{equation}
	We define linear polarization four-vectors
	\begin{equation}
		\epsilon_1^\mu = (0,\epsilon_1^-, 1, 0) \qquad \epsilon_2^\mu = (0,\epsilon_2^-, 0, 1).
	\end{equation}
	In the free limit $e=0$ the constraint equation determines $\epsilon_j^- = \frac{2k^j}{k^+}$, and the photon is transversely polarized, $\epsilon_j^\mu k_\mu=0$.

	With these definitions, the Schr\"{o}dinger picture mode expansions are
	\begin{equation}
		\psi_- = \int \frac{dp^+ dp^\perp}{(2\pi)^3} \sum_{s=\pm\frac{1}{2}} e^{-i\mathsf{p}\mathsf{x}} w^s a^s_p + e^{i\mathsf{p}\mathsf{x}} w^{-s} b^{s\dag}_p \qquad A^j = \int \frac{dk^+ dk^\perp}{(2\pi)^3\sqrt{k^+}} \left( e^{-i\mathsf{k}\mathsf{x}} c^j_k + e^{i\mathsf{k}\mathsf{x}} c^{j\dag}_k \right).
		\label{fields_exp}
	\end{equation}
	These fields can then be substituted into the Hamiltonian densities to obtain the  Schr\"odinger-picture Hamiltonian:
	\begin{equation}
		H = \frac{1}{2} \int \mathcal{H} \ d^3\mathsf{x}.
	\end{equation}
	Subsequent integrals may be evaluated by using
	\begin{equation}
		\int d^3\mathsf{x} \ e^{i(\mathsf{p} - \mathsf{p}')\mathsf{x}} = 2(2\pi)^3 \delta^3(\mathsf{p}-\mathsf{p}').
		\label{delta}
	\end{equation}
	
	After constructing the Hamiltonian in this way one takes the additional step of normal-ordering the creation and annihilation operators. (See App. \ref{app_A} for sample calculations.) This procedure renormalizes away the simplest ``ear diagram" divergences, and it clarifies that the Fock vacuum is the exact ground state in the light-front. 
	
	Simulating the theory requires discretizing the momenta, truncating the set of single-particle states, and truncating the photon state occupation numbers. We use a momentum lattice with lattice spacing $\frac{2\pi}{L}$. The creation and annihilation operators then obey discrete (anti)commutation relations:
	\begin{equation}
		\{ a^s_p, a^{r\dag}_{p'} \} = \delta_{s,r} \delta_{\mathsf{p}, \mathsf{p}'} \qquad \{ b^s_p, b^{r\dag}_{p'} \} = \delta_{s,r} \delta_{\mathsf{p}, \mathsf{p}'} \qquad [ c^j_k, c^{j'\dag}_{k'} ] = \delta_{j,j'} \delta_{\mathsf{k}, \mathsf{k}'}.
		\label{discrete_commutators}
	\end{equation}
	As a result, they are scaled by a factor of $\sqrt{L^3}$ relative to the infinite volume case. For example, $a_p^s \to \sqrt{L^3} a_p^s$ in passing from the continuum to discretized single-particle state space.
	
	\subsection{Interaction Picture}
	
	The SFQED Schr\"odinger-picture Hamiltonian can be split into a ``free" piece, quadratic in the fluctuating fields, and an ``interaction" piece, accounting for interactions of order $e$ and $e^2$. The free Hamiltonian $H_0 = H_\psi + H_A + H_{\psi \mathcal{A}}$ is composed of number operators, so it is  diagonal in the Fock basis, whereas the interaction Hamiltonian $V = H_{\psi A \mathcal{A}} + H_{\psi A} + H_{4\psi} + H_{2\psi A}$ is not. As a result it is convenient to work in an interaction picture. Since even the Schr\"odinger-picture Hamiltonian is time-dependent in the present context, let us review the derivation of the interaction picture time evolution operator.
	
	The Schr\"odinger-picture time evolution operator is 
	\begin{align}
		U = 
		\mathcal{T} e^{-\frac{i}{2}\int_0^{x^+} H_{S}(y^+) dy^+}
		\label{schro_time_evo}
	\end{align}
	where $H_S(x^+)$ is the complete time-dependent Schr\"odinger-picture Hamiltonian. The lower limit of integration is set to zero but in general denotes some initial fiducial time at which the Schr\"odinger-, interaction-, and Heisenberg-picture states are the same. 
	Schr\"odinger-picture states are evolved with $U$, $|\psi_S(x^+)\rangle = U|\psi(0)\rangle$, and satisfy the Schr\"odinger equation with $H_S$. Now define the free evolution operator
	\begin{align}
		U_0 = 
		\mathcal{T} e^{-\frac{i}{2}\int_0^{x^+} H_{0}(y^+) dy^+}
		\label{schro_time_evo0}
	\end{align}
	and the interaction picture states
	\begin{align}
		|\psi_{int}(x^+)\rangle &= U_0^\dagger |\psi_S(x^+)\rangle \nonumber\\
		&= U_0^\dagger U|\psi(0)\rangle\nonumber\\
		&= U_{int}(x^+)\ket{\psi(0)}.
	\end{align}
	In the last line we have defined the  interaction-picture time evolution operator $U_{int}=U_0^\dagger U$. It can be shown to satisfy
	\begin{align}
		i \partial_+ U_{int} &= \frac{1}{2} (U_0^\dagger V U_0 )U_{int}\nonumber\\
		&= \frac{1}{2} H_{int} U_{int}
		\label{interactionpicschroeq}
	\end{align}
	where the second line defines the interaction-picture interaction Hamiltonian $H_{int}$. (Essentially it amounts to inserting the Volkov mode solutions~\cite{volkov_1935} into the Schr\"odinger-picture interaction Hamiltonian $V$ -- see Appendix \ref{app_A}.)  The solution to Eq.~(\ref{interactionpicschroeq}) is 
	\begin{align}
		U_{int} = \mathcal{T} e^{-\frac{i}{2}\int_{0}^{x^+} H_{int}(y^+) dy^+}.
		\label{time_evo}
	\end{align}
	The Schr\"odinger picture time evolution operator can be written as $U = U_0 U_{int}$, where $U_0$ is entirely diagonal and amounts to phases when acting on free-particle basis states, so the typical probability between these states can be calculated just using $U_{int}$. 
	
	\section{Quantum Circuit Design}
	\label{sec_quantum_circuit_design}
	
\begin{figure}[h]
		\centering
		\begin{tikzpicture}
			\begin{feynman}
				\vertex (a) {$\gamma$};
				\vertex [right=of a] (b);
				\vertex [above right=of b] (f1) {$e^-$};
				\vertex [below right=of b] (f2) {$e^+$};
				\diagram*{
					(a) -- [boson] (b),
					(b) -- [\volkovprop] (f1),
					(f2) -- [\volkovprop] (b)
				};
			\end{feynman}
		\end{tikzpicture}
		\caption{Nonlinear Breit-Wheeler pair-production. Double lines indicate $e^+e^-$ propagation on the background field. This ``decay" proceeds at tree level in SFQED in the Furry expansion \cite{PhysRev.81.115} (see Eq.~(\ref{bw_int_hamiltonian})). Applying the Jordan-Wigner transformation to represent three one-particle states by  qubits, we translate the time evolution operator into a quantum circuit.}
		\label{breit_wheeler}
	\end{figure}
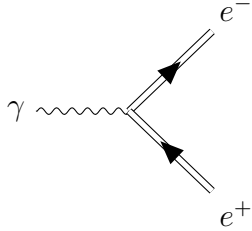

	There are many ways to build the same unitary as a quantum circuit, but in the NISQ era we must be careful to minimize the use of noisy gates, particularly CNOTs and SWAPs. At present we cannot afford to simulate the most scientifically interesting SFQED processes with many degrees of freedom. However, as a first step toward the long-term goal and as a proof of principle, we will show that a simple example process, a tree-level three-body process with exclusive final states, is reliably simulated. In this section we build three-qubit circuits that optimally simulate the parts of the SFQED Hamiltonian responsible for the tree-level nonlinear Breit-Wheeler process, shown in Fig.~\ref{breit_wheeler}.
	
	We encode each particle in a qubit $q$. For the processes considered here it is sufficient to truncate boson occupation numbers at one. If $q$ is in the $\ket{0}$ state, then the Fock state is unoccupied; if $q$ is in the $\ket{1}$ state, the Fock state is occupied. For a truncated Hilbert space built from one single-particle state each for an electron, positron, and photon, the encoding is given by a bitstring of three qubits, such as $\ket{001}$, written as a Fock state according to Eq.~(\ref{fock}). Thus the state $\ket{001}$ is a single photon and $\ket{110}$ is an electron-positron pair. We follow Qiskit's little-endian notation, so that the qubits are numbered $\ket{q_2,q_1,q_0}$. That is, the photon is qubit zero, $q_0$.
	
	The terms in the Hamiltonian that describe the transition between these states are those with the operators $a^\dag b^\dag c$ and $c^\dag ba$. We transcribe these two operator monomials  into quantum gates with the Jordan-Wigner transformation. For example:
	\begin{equation}
		b = \frac{1}{2} (Z_2) (X_1+iY_1) \qquad c^\dag = \frac{1}{2}(X_0-iY_0),
	\end{equation}
	where $X_i$, $Y_i$, and $Z_i$ are the Pauli operators acting on the $i$th qubit.  $Z$ gates account for the fermion $\zeta$ factor. By using Pauli identities, we can write the monomials as a linear combination of Pauli strings, each written purely with $X$ and $Y$ gates. In the case of three qubits, there are $2^3=8$ different terms in the linear combination, since each qubit can be acted on with either an $X$ gate or a $Y$ gate. In other words,
	\begin{IEEEeqnarray*}{rCl}
		a^\dag b^\dag c & \sim & c_{XXX}XXX + c_{XXY}XXY + c_{XYX}XYX + c_{XYY}XYY \\
		& + & c_{YXX}YXX + c_{YXY}YXY + c_{YYX}YYX + c_{YYY}YYY. \yesnumber
		\label{paulis}
	\end{IEEEeqnarray*}
	
	We emphasize that because the interactions in the Hamiltonian are limited to three- and four-body,  the Pauli string scaling is  not exponential but rather polynomial in the number of qubits. (The nonlocality in momentum space increases the connectivity and therefore the power of the polynomial; however, by solving the constraints explicitly, even a position-space lattice implementation would have higher connectivity.)
	
	Although the true time evolution operator is given in Eq. (\ref{time_evo}), this operator must be discretized as well to digitally simulate time evolution. The typical prescription is to use the Lie-Trotter approximation, also referred to as first-order Trotterization. Although higher order Trotterizations such as the second-order Suzuki-Trotter formula lead to better accuracy for longer time steps, the circuit length is increased per time step, leading to more quantum error. It is a problem-dependent question whether it is worthwhile to implement the second-order formula, and in our case, we found that first-order is preferable. Additionally, we wish to see fine-grained details not found using long time steps. Therefore we write
	\begin{equation}
		U_{int}(x^+) \approx e^{-\frac{i}{2}H_{int}(x_n^+)\Delta x^+} \cdots e^{-\frac{i}{2}H_{int}(x_1^+)\Delta x^+},
		\label{u_trotter}
	\end{equation}
	where $n$ denotes the number of time steps used in the approximation, so $\frac{x^+}{n} = \Delta x^+$ and $x_n^+ = n\Delta x^+$.
	
	As a result, we must build a quantum circuit representing the matrix exponential of Eq.~(\ref{paulis}).  Here again we opt for a first-order as opposed to second-order Trotterization for the same reasons discussed above, so that
	\begin{IEEEeqnarray*}{rCl}
		e^{-\frac{i}{2}\Delta x^+ a^\dag b^\dag c} & \sim & e^{-\frac{i}{2}\Delta x^+ c_{XXX}XXX} e^{-\frac{i}{2}\Delta x^+ c_{XXY}XXY} e^{-\frac{i}{2}\Delta x^+ c_{XYX}XYX} e^{-\frac{i}{2}\Delta x^+ c_{XYY}XYY} \\
		& \times & e^{-\frac{i}{2}\Delta x^+ c_{YXX}YXX} e^{-\frac{i}{2}\Delta x^+ c_{YXY}YXY} e^{-\frac{i}{2}\Delta x^+ c_{YYX}YYX} e^{-\frac{i}{2}\Delta x^+ c_{YYY}YYY}. \yesnumber
		\label{trotter}
	\end{IEEEeqnarray*}
	Each exponential factor now forms a subcircuit. Importantly, the order of these factors does not matter in the approximation, and it is very useful to shift certain subcircuits around and apply quantum gate identities to minimize the number of gates, particularly CNOTs.
	
\begin{figure*}[ht]
		\begin{center}
			$S_X \equiv $ \begin{quantikz}
				\lstick{$q_0$} & & \targ{} & & \\
				\lstick{$q_1$} & \ctrl{1} & \ctrl{-1} & \gate{H} & \\
				\lstick{$q_2$} & \targ{} & & &
			\end{quantikz} \hspace{1cm}
			$S_Y \equiv $ \begin{quantikz}
				\lstick{$q_0$} & & \targ{} & & & \\
				\lstick{$q_1$} & \ctrl{1} & \ctrl{-1} & \gate{S} & \gate{H} & \\
				\lstick{$q_2$} & \targ{} & & & &
			\end{quantikz}
		\end{center}
		\caption{GHZ diagonalization circuits for Pauli strings with an odd number of $X$s (left) or odd number of $Y$s (right). Diagonalization facilitates implementation of the  time evolution subcircuits with minimal gate counts.}
		\label{ghz}
	\end{figure*}
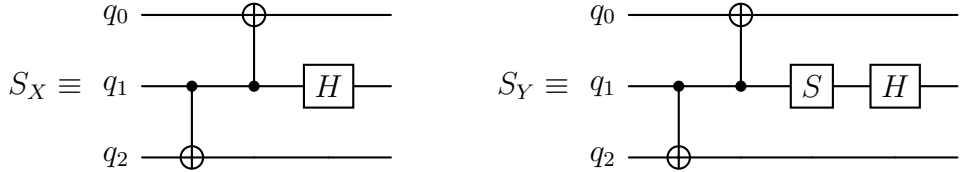

	There are numerous ways to simulate each subcircuit. We employ the GHZ diagonalization used in previous literature~\cite{Stetina2022simulatingeffective, PhysRevD.107.054512}. Subcircuits for Pauli strings with an odd number of $X$ ($Y$) gates can be diagonalized with a GHZ transformation circuit.  The transformation circuits are shown in Fig.~\ref{ghz} and the diagonalization relations are shown in Tab.~\ref{diagonalization}. We note that the qubit that is acted on by $X$- or $Y$-basis transformation gates is the qubit that will always have a $Z$ gate upon diagonalization. Therefore, to reduce the use of long-ranged CNOT gates, it is ideal to place these transformation gates at a middle qubit, making the middle qubit the control qubit.
	
	\begin{table}[ht]
		\begin{center}
			\begin{tabular}{c|c||c|c}
				Odd $X$s & $S_X (\text{string}) S_X^\dag$ & Odd $Y$s & $S_Y (\text{string}) S_Y^\dag$ \\ \hline
				$XXX$ & $IZI$ & $YYY$ & $ZZZ$ \\ \hline
				$YYX$ & $-ZZI$ & $XXY$ & $-IZZ$ \\ \hline
				$YXY$ & $-ZZZ$ & $XYX$ & $-IZI$ \\ \hline
				$XYY$ & $-IZZ$ & $YXX$ & $-ZZI$
			\end{tabular}
		\end{center}
		\caption{ Diagonalizations of Pauli strings consisting of $X$ and $Y$ operators by the operators of Fig.~\ref{ghz}. }
		\label{diagonalization}
	\end{table}
	
	Since $S_X$ and $S_Y$ are unitary, we may write $e^{S_X^\dag A S_X} = S_X^\dag e^A S_X$. Then by rearranging the matrix exponentials in Eq.~(\ref{trotter}), we can write the Trotterization with one half diagonalized by $S_X$ and the other half diagonalized by $S_Y$. This way Eq.~(\ref{trotter}) becomes
	\begin{IEEEeqnarray*}{rCl}
		e^{-\frac{i}{2}\Delta x^+ a^\dag b^\dag c} & \sim & S_X^\dag e^{-\frac{i}{2}\Delta x^+ c_{XXX}IZI} e^{\frac{i}{2}\Delta x^+ c_{YYX}ZZI} e^{\frac{i}{2}\Delta x^+ c_{YXY}ZZZ} e^{\frac{i}{2}\Delta x^+ c_{XYY}IZZ} S_X \\
		& \times & S_Y^\dag e^{-\frac{i}{2}\Delta x^+ c_{YYY}ZZZ} e^{\frac{i}{2}\Delta x^+ c_{XXY}IZZ} e^{\frac{i}{2}\Delta x^+ c_{XYX}IZI} e^{\frac{i}{2}\Delta x^+ c_{YXX}ZZI} S_Y. \yesnumber
	\end{IEEEeqnarray*}
	Each matrix exponential can now be simulated solely with CNOTs and $R_Z$ gates. The four relevant circuits are shown in Fig. \ref{diags} and the final circuit is shown in Fig. \ref{circuit}.
	
	\begin{figure}[ht]
		\begin{center}
			$e^{-\frac{i\theta}{2}IZI} = $ \begin{quantikz}
				\lstick{$q_0$} & & \\
				\lstick{$q_1$} & \gate{R_Z(\theta)} & \\
				\lstick{$q_2$} & &
			\end{quantikz} \hspace{1cm}
			$e^{-\frac{i\theta}{2}ZZZ} = $ \begin{quantikz}
				\lstick{$q_0$} & \ctrl{1} & & & & \ctrl{1} & \\
				\lstick{$q_1$} & \targ{} & \ctrl{1} & & \ctrl{1} & \targ{} & \\
				\lstick{$q_2$} & & \targ{} & \gate{R_Z(\theta)} & \targ{} & &
			\end{quantikz}
			
			\vspace{1cm}
			
			$e^{-\frac{i\theta}{2}ZZI} = $ \begin{quantikz}
				\lstick{$q_0$} & \ctrl{1} & & \ctrl{1} & \\
				\lstick{$q_1$} & \targ{} & \gate{R_Z(\theta)} & \targ{} & \\
				\lstick{$q_2$} & & & &
			\end{quantikz} \hspace{1cm}
			$e^{-\frac{i\theta}{2}IZZ} = $ \begin{quantikz}
				\lstick{$q_0$} & & & & \\
				\lstick{$q_1$} & \ctrl{1} & & \ctrl{1} & \\
				\lstick{$q_2$} & \targ{} & \gate{R_Z(\theta)} & \targ{} &
			\end{quantikz}
		\end{center}
		\caption{Time evolution subcircuits after  diagonalization. In Qiskit's convention $R_Z(\theta) \equiv e^{-\frac{i\theta}{2}Z}$. The CNOT gates in each subcircuit compute/uncompute the parity of the incoming bitstring.}
		\label{diags}
	\end{figure}
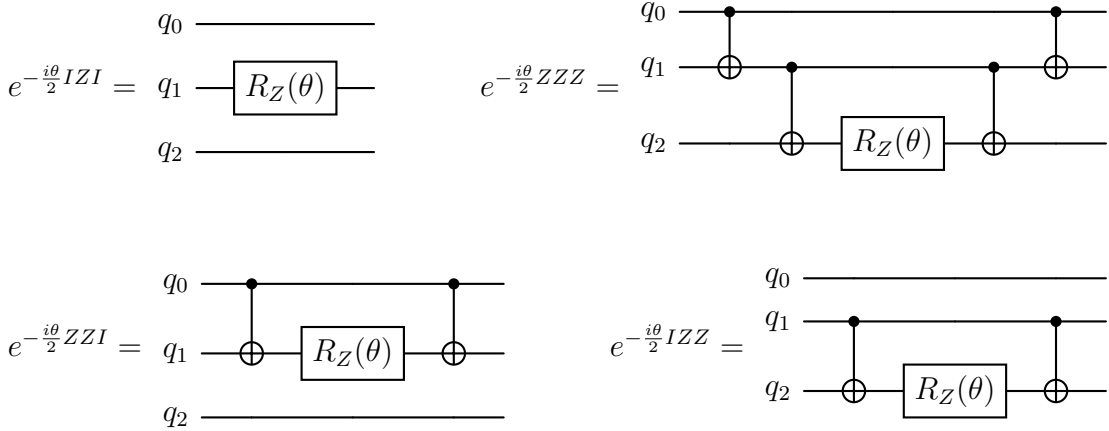
	
	\begin{sidewaysfigure}
		$e^{-\frac{i}{2}H(x_i^+) \Delta x^+} \approx $
		\begin{tikzpicture}
			\node[scale=0.667]{
				\begin{quantikz}
					\lstick{$q_0$} & & \targ{} & & \ctrl{1} & & & & \ctrl{1} & & & & & & & & \ctrl{1} & & & & \ctrl{1} & & & \targ{} & & \\
					\lstick{$q_1$} & \ctrl{1} & \ctrl{-1} & \gate{H} & \targ{} & \ctrl{1} & \gate{R_Z(-c_{XYY})} & \ctrl{1} & \targ{} & \ctrl{1} & \gate{R_Z(c_{XXX})} & \ctrl{1} & \gate{\sqrt{X}} & \ctrl{1} & \gate{R_Z(-c_{XXY})} & \ctrl{1} & \targ{} & \ctrl{1} & \gate{R_Z(-c_{YXX})} & \ctrl{1} & \targ{} & \gate{H} & \gate{S^\dag} & \ctrl{-1} & \ctrl{1} & \\
					\lstick{$q_2$} & \targ{} & & & & \targ{} & \gate{R_Z(-c_{YXY})} & \targ{} & & \targ{} & \gate{R_Z(-c_{YYX})} & \targ{} & & \targ{} & \gate{R_Z(-c_{XYX})} & \targ{} & & \targ{} & \gate{R_Z(c_{YYY})} & \targ{} & & & & & \targ{} &
				\end{quantikz}
			};
		\end{tikzpicture}
		\caption{First-order Trotterization circuit for a single time evolution step. By alternating the $S_X$ and $S_Y$ halves of the circuit in subsequent time steps, cancellations can be used to minimize the gate count. The first step has 16 CNOT gates, but only 12 CNOT gates are added with each succeeding step.}
		\label{circuit}
	\end{sidewaysfigure}
	
	In practice, IBM quantum computers use a limited set of native quantum gates: $X$, $R_Z(\theta)$, CNOT, $I$, and $\sqrt{X}$. Thus, the final step in circuit creation is to translate the circuit into these gates. For example, the Hadamard gate becomes $H = R_Z(\frac{\pi}{2}) \sqrt{X} R_Z(\frac{\pi}{2})$ with an overall phase of $e^{i\frac{\pi}{4}}$, which falls out of the probability.
	
	\section{Nonlinear Breit-Wheeler and the null double slit experiment}
	\label{sec_breit_wheeler}
	
	In the previous section we constructed a gate implementation for a truncation of the SFQED Hamiltonian. The truncation was drastic, but designed to  capture the tree-level Breit-Wheeler interaction of Fig.~\ref{breit_wheeler} with fixed  arbitrary initial and final states, and arbitrary null background fields. With this simple first-order interaction, it is possible to realize interesting quantum dynamics. An elegant example is a null version of the double-slit interference pattern~\cite{Ilderton_2020, Ilderton_2019}. We will use this process as a benchmark for quantum simulation of the truncated SFQED Hamiltonian, computing the real-time probabilities and comparing with classical simulation. 
	
	First, to build intuition, we describe the process using first order perturbation theory. Actually, although the truncated Hamiltonian operates on an eight dimensional Fock space, for the purposes of the Breit-Wheeler process, we need only consider a two-level subsystem corresponding to the basis states $\ket{e^-e^+} = (1,0)$ and $\ket{\gamma} = (0,1)$. We are interested in the amplitude $\bra{e^-e^+}U\ket{\gamma}$.
	
	The simplest Hamiltonian truncation corresponds to the Hilbert subspace where the electron and positron have the same helicity with equal and opposite transverse momenta. Let the helicities be $s=\frac{1}{2}$, and let the photon polarization be $\lambda=1$. We take the photon momentum to be $K^\mu = (2p^+, 0, 0, 0)$, the electron momentum to be $P^\mu = (p^+, p^-, p^1, 0)$, and the positron momentum to be $Q^\mu = (p^+, p^-, -p^1, 0)$. In the latter two cases the mass shell condition fixes $p^-=\frac{(p^1)^2+m^2}{p^+}$.
	
\begin{figure}
		\centering
		\includegraphics[scale=0.75]{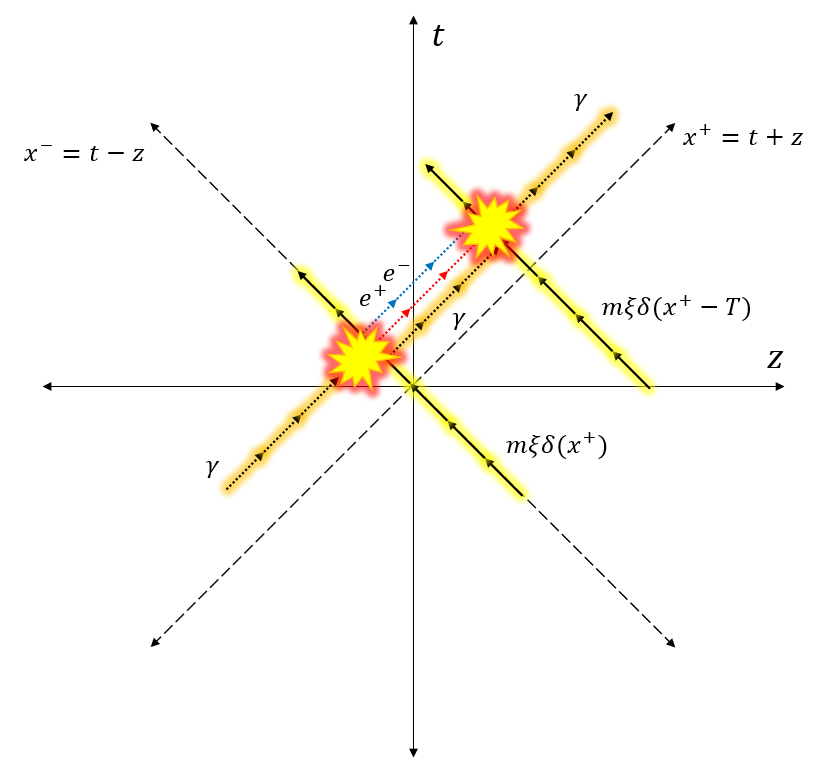}
		\caption{Two delta-function laser pulses propagate in the $-\hat{z}$ direction, separated in time. A photon collides head-on with the first pulse, leading to a superposition of the $\ket{e^-e^+}$ and $\ket{\gamma}$ states. Collision with the second pulse may cause attenuation or enhancement of the $\ket{e^-e^+}$ state, depending on the pulse separation. Sketched is a case of total destructive interference in pair-production probability for a given electron-positron momentum mode. Light-front coordinates are convenient because all particles interact with the laser pulses at the same light-front time $x^+$.}
		\label{spacetime_diagram}
	\end{figure}

	The background field we will consider corresponds to a linearly-polarized plane wave $\mathcal{A}^\mu(\kappa_\mu x^\mu) = (0,0,\mathcal{A}^1(\kappa_\mu x^\mu), 0)$.  For a null field traveling in the $-\hat{z}$-direction, we have $\kappa_\mu x^\mu = \omega x^+$. The electric field strength $E$ can be nondimensionalized as $\xi = \frac{eE}{m\omega}$, where $e$ is the gauge coupling \cite{PhysRevD.60.092004}. A simple case to consider is a very short-duration pulse. An example is given by the gauge potential
	\begin{equation}
		e\mathcal{A}^1(\omega x^+) = m\xi(1+\tanh(\omega x^+)).
	\end{equation}
	Analytical treatment becomes  straightforward if we take the limit of $\omega \to \infty$, creating a delta-function pulse and a Heaviside-function potential:
	\begin{equation}
		\lim\limits_{\omega \to \infty} e\mathcal{A}^1(\omega x^+) = m\xi \theta(x^+).
	\end{equation}
	In the null double-slit process of~\cite{Ilderton_2020}, we allow the photon to collide with two such pulses separated in light-front time.  Let one pulse appear at $x^+=0$ and the other appear at $x^+=T$. Then our background field is
	\begin{equation}
		e\mathcal{A}^1(x^+) = m\xi[\theta(x^+) + \theta(x^+ - T)].
	\end{equation}
	Fig. \ref{spacetime_diagram} shows a spacetime diagram of this nonlinear Breit-Wheeler process.
	
	\subsection{Perturbation theory}
	
	Since the process is relatively  simple, we can describe the dynamics with first-order perturbation theory. This analysis is somewhat tangent to the main purpose of the paper, for which we could make do merely with a quantum simulation and a classical exact diagonalization numerical treatment for comparison. However, it is helpful to have an analytic description in order to understand the physics of the results and the limitations of perturbation theory.
	
	To calculate the probability $|\bra{e^-e^+} U(x^+) \ket{\gamma}|^2$ we expand the interaction-picture time evolution operator to first order,
	\begin{equation}
		U(x^+) \rightarrow 1 - \frac{i}{2} \int_{0}^{x^+} H_{int}(y^+) \ dy^+.
		\label{perturb_approx}
	\end{equation} 
	The interaction-picture Hamiltonian consists of
	\begin{equation}
		H_{int}(y^+) = - \frac{2me}{\sqrt{2{p^+}^3 L^3}} e^{i p^- y^+} e^{i(f(P)+g(Q))} a^{+\frac{1}{2}\dag}_P b^{+\frac{1}{2}\dag}_Q  c^{+1}_K.
	\end{equation}
	Here
	\begin{equation}
		f(P) + g(Q) = \int_{0}^{y^+} \frac{dz^+}{p^+} \left[ -2p^1 e \mathcal{A}^1(z^+) + (e\mathcal{A}^1(z^+))^2 \right].
	\end{equation}
	Using $\int_{0}^{y^+} \theta(z^+-T)^n dz^+ = (y^+-T) \theta(y^+-T)$, we can evaluate the phase to be
	\begin{equation}
		f(P) + g(Q) = \frac{m\xi}{p^+}[ (m\xi-2p^1) y^+ \theta(y^+) + (3m\xi - 2p^1)(y^+-T)\theta(y^+-T) ].
	\end{equation}
	Now let us define the parameter combinations 
	\begin{align}
		\alpha &= \frac{m\xi(m\xi-2p^1)}{p^+},\nonumber\\
		\beta &= \frac{m\xi(3m\xi-2p^1)}{p^+}.
	\end{align}
	Rewriting $e^{a \theta(y^+)}=(e^a - 1)\theta(y^+) + 1$, $e^{i(f(P)+g(Q))}$ is expressed compactly as
	\begin{equation}
		e^{i(f(p)+g(q))} = e^{i\alpha y^+}(1-\theta(y^+-T)) + e^{i(\alpha+\beta)y^+} e^{-i\beta T} \theta(y^+-T).
	\end{equation}
	Note that $\theta(y^+)\rightarrow 1$ since we are considering $y^+ \geq 0$. Putting together the pieces we find
	\begin{equation}
		\bra{e^-e^+}H_{int}(y^+)\ket{\gamma} = \frac{2me}{\sqrt{2{p^+}^3L^3}} \left[ e^{i(\alpha+p^-)y^+}(\theta(y^+-T)-1) - e^{i(\alpha+\beta+p^-)y^+}e^{-i\beta T}\theta(y^+-T) \right].
	\end{equation}
	Integrating over light-front time, we have the probability amplitude 
	\begin{IEEEeqnarray*}{rCl}
		\bra{e^-e^+}U(x^+)\ket{\gamma} & \approx & \frac{me}{\sqrt{2{p^+}^3L^3}} \left[ \frac{e^{i(\alpha+p^-)x^+}}{\alpha+p^-} (1 - \theta(x^+-T)) - \frac{1}{\alpha+p^-} \right. \\
		& + & \frac{e^{i(\alpha+p^-)T}}{\alpha+p^-} \theta(x^+-T) + \frac{e^{i(\alpha+\beta+p^-)x^+}}{\alpha+\beta+p^-} e^{-i\beta T}\theta(x^+-T) \\
		& - &\left. \frac{e^{i(\alpha+p^-)T}}{\alpha+\beta+p^-} \theta(x^+-T) \right]. \yesnumber
	\end{IEEEeqnarray*}
	We can find separate probabilities for $x^+ \gtrless T$. If $x^+ < T$, then the probability of pair-production is
	\begin{equation}
		P_{\gamma \to e^-e^+}(x^+ < T) = \frac{2m^2e^2}{ (\alpha+p^-)^2 {p^+}^3 L^3} \sin^2\left( \frac{\alpha+p^-}{2} x^+ \right).
		\label{lessT}
	\end{equation}
	If $x^+ > T$, then the probability is
	\begin{IEEEeqnarray*}{rCl}
		P_{\gamma \to e^-e^+}(x^+ > T) & = & \frac{m^2e^2}{(\alpha+p^-)^2(\alpha+\beta+p^-)^2{p^+}^3 L^3} \left[ \alpha^2 + \beta^2 + {p^-}^2 + \alpha(\beta+2p^-) + \beta p^- \right. \\
		& - & \beta(\alpha+\beta+p^-)\cos[(\alpha+p^-)T] + \beta(\alpha+p^-)\cos[(\alpha+\beta+p^-)(x^+-T)] \\
		& - & \left. (\alpha+p^-)(\alpha+\beta+p^-)\cos[(\alpha+\beta+p^-)x^+ - \beta T] \right]. \yesnumber
		\label{moreT}
	\end{IEEEeqnarray*}
	
	Let us now inspect Eq. (\ref{lessT}), the probability after the first pulse has arrived.  Time-averaging over a cycle, the probability is 
	\begin{equation}
		\expect{P}_{\gamma \to e^-e^+}(x^+ < T) = \frac{m^2e^2}{(\alpha+p^-)^2{p^+}^3 L^3},
	\end{equation}
	which is maximized with respect to the background field strength when $\alpha$ is minimized. This corresponds to $m\xi = p^1$, or $\alpha = -\frac{(p^1)^2}{p^+}$ and $\beta = -\alpha$. Consequently, if we tune $p^1$ near to $m\xi$ we maximize the conversion probability. Furthermore, we note that since $\alpha$ is parabolic in $m\xi$, there will always be pairs of momenta modes that reach the same level of enhancement.
	
	Restoring the oscillating part of Eq. (\ref{lessT}), we also see that the maximum probability is achieved when $x^+ = \frac{(2n+1)\pi}{\alpha + p^-}$ for $n \in \mathbb{Z}^+$. 
	
	Let us now inspect Eq. (\ref{moreT}). Taking $\beta=-\alpha$ and time-averaging, we find the following ratio with Eq. (\ref{lessT})
	\begin{equation}
		\frac{\expect{P}_{\gamma \to e^-e^+}(x^+ > T)}{\expect{P}_{\gamma \to e^-e^+}(x^+ < T)} = 1 + \frac{\alpha^2}{(p^-)^2} + \frac{2\alpha}{p^-} \cos^2\left( \frac{\alpha + p^-}{2}T \right).
	\end{equation}
	To maximize this ratio, we recognize that $\alpha < 0$, so we need $T = \frac{(2n+1)\pi}{\alpha + p^-}$. If we further take the high-energy limit $p^1 \gg m$, then $\alpha = -p^-$ and we obtain
	\begin{equation}
		\frac{\expect{P}_{\gamma \to e^-e^+}(x^+ > T)}{\expect{P}_{\gamma \to e^-e^+}(x^+ < T)} = 2.
	\end{equation}
	This gives the factor of enhancement from the first time-averaged probability to the next.
	
	\subsection{Classical Simulation}
	
	In the near-term the accuracy of quantum computations must be benchmarked against classical simulations.We begin by configuring the momentum lattice. The lattice spacing is $\frac{2\pi}{L}$ and to approximate the continuum limit, we need this spacing to be less than the typical momentum in the processes of interest. We will take a large $L=\frac{50\pi}{m}$.
	
	Let us now investigate the pair-production of electron-positron pairs with differing transverse momenta. As discussed above the electron four-momentum is given by $P^\mu = (p^+, p^-, p^1, 0)$, the positron four-momentum is $Q^\mu = (q^+, q^-, q^1,0)$, and the photon four-momentum $K^\mu = (k^+, 0, 0, 0)$. Here $p^+=q^+$, so $k^+=2p^+$, and $q^1=-p^1$. Due to equal and opposite transverse momenta, we can classify an electron-positron pair by the produced electron transverse momentum $p^1$.
	
	For concreteness we will take $\xi = 6$, so that the maximum possible enhancement is seen for $p^1 = 6m$. In addition to simulating this momentum mode, we simulate also $p^1=0$ and $p^1=12m$ modes, which should have equal enhancements to each other. $p^+$ is arbitrary and we fix it to $7.2m$. Correspondingly, the light-front energies are $k^-=0$ for the photon, and $p^-+q^-\approx (0.3,10,40)m$ for $p^1=(0,6,12)m$, respectively. Physically, the sign of the momentum results from a negative transverse electric field $E^\perp=-\partial_+ \mathcal{A}^\perp$. Positive particles will experience a force in the $-\hat{x}^1$ direction while negative particles will be accelerated in the $+\hat{x}^1$ direction.
	
	In addition to momentum, the Fock states possess helicity and/or polarization quantum numbers. We let the initial photon be an equal superposition of both polarizations. We will simulate the production of electron-positron pairs with all four different helicity configurations, and sum their probabilities to obtain the final pair-production probability. Accordingly, we keep two photon polarization states for a single momentum mode and four electron-positron helicity states for three electron-positron momentum modes. This creates a 14-dimensional Hilbert space.
	
	Lastly, let us discuss state preparation. The Fock states are not eigenstates of the interacting Hamiltonian. During time evolution, Fock states evolve into superpositions of the true eigenstates. We use adiabatic state preparation to map the initial and final states back and forth between these bases, adiabatically turning on/off the electric coupling $e$. Adiabatic turn-on/off is performed linearly over 1000 time units, compared to the physical evolution time of 100 units during the two-pulse encounter, where the coupling is constant. However, the effects of adiabatic turn-on are  negligible for the finite $p^1$ modes due to kinematic suppression (these modes are of high light-front energy). Since adding thousands of Trotter steps is prohibitive in the NISQ era, we will focus on the $p^1=6m$ mode in the quantum simulations, for which adiabatic turn-on can be neglected\footnote{This remains true even with the artificially inflated value of the coupling used in the quantum simulation, since mixing with the incoming photon state is still strongly kinematically suppressed.}. Furthermore we measure probabilities just after the collision with the second pulse, neglecting adiabatic turn-off, which is also a good approximation for this mode.
	
	\begin{figure}
		\centering
		\begin{subfigure}{\linewidth}
			\centering
			\includegraphics[scale=0.42]{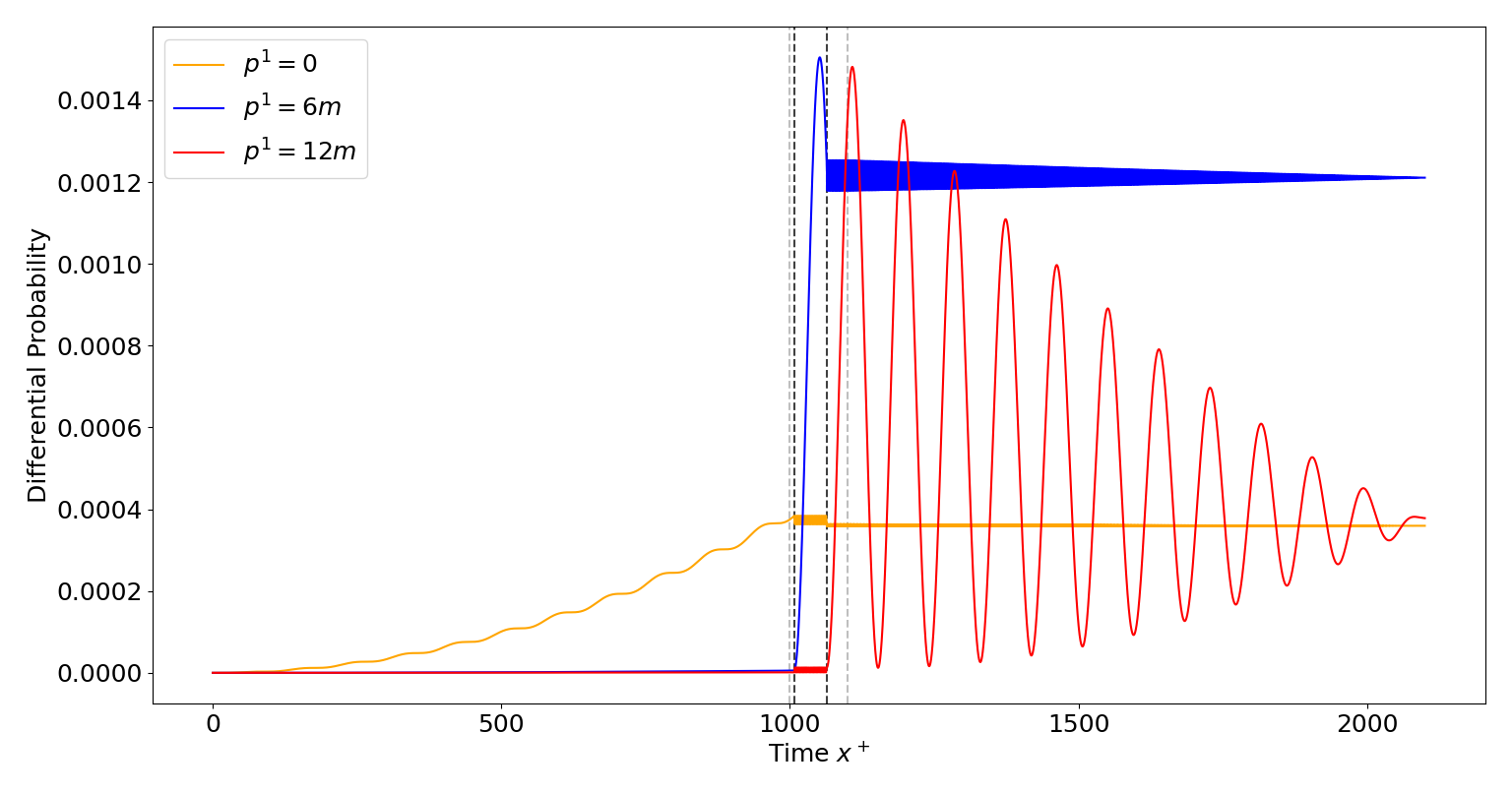}  
			\label{theta_1}
		\end{subfigure}
		
		\begin{subfigure}{\linewidth}
			\centering
			\includegraphics[scale=0.42]{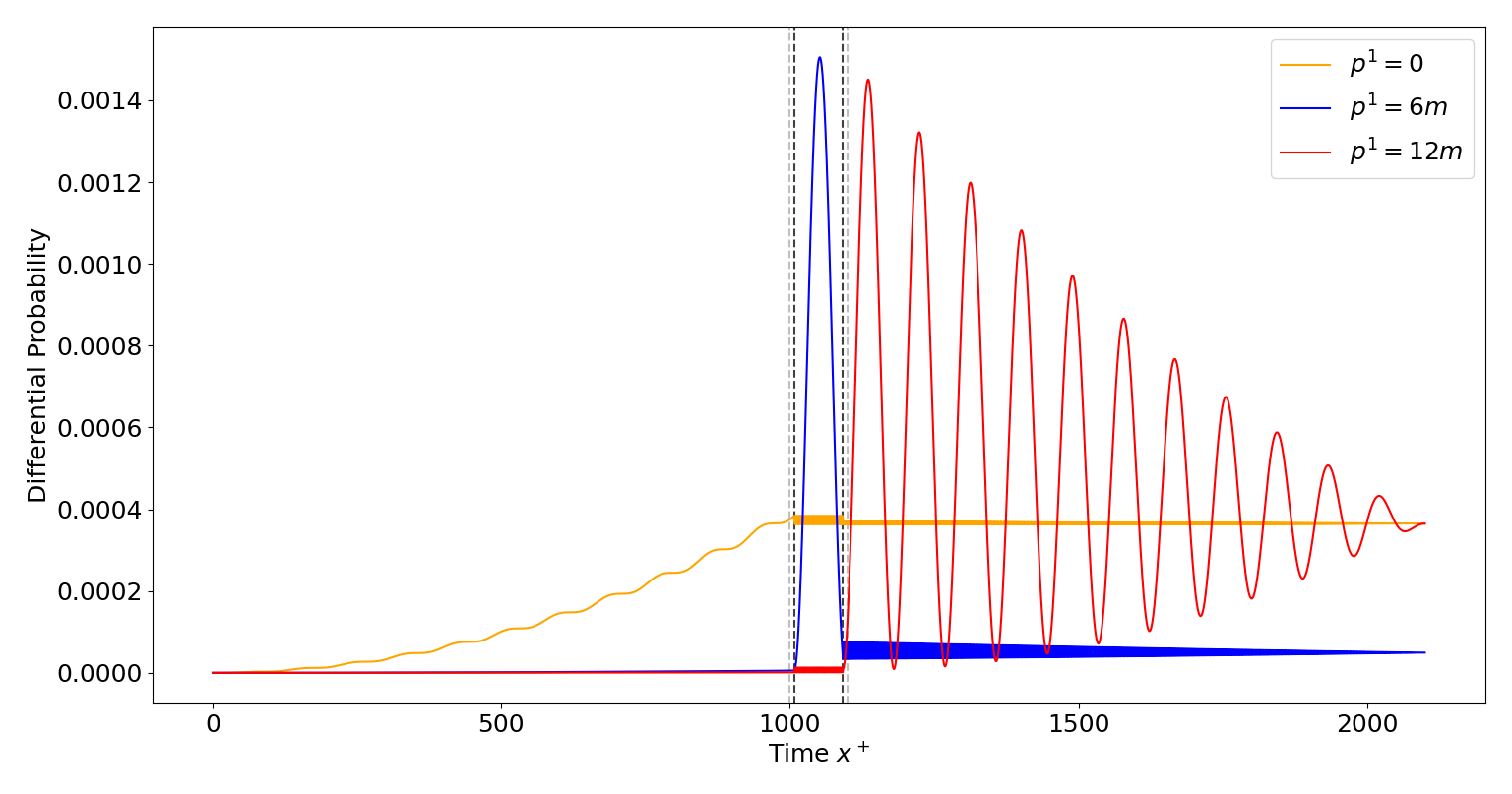}  
			\label{theta_1_5}
		\end{subfigure}
		\caption{Classical simulations of nonlinear Breit-Wheeler pair-production differential probabilities as a function of light-front time. The three curves represent three different electron-positron pairs, labeled by the $p^1$ electron transverse momentum. The vertical light gray lines enclose the physical evolution period when the coupling is held constant. Each dark gray line represents a collision of the incoming photon with the background laser pulse. Only the $p^1=m\xi$ mode sees significant constructive or destructive interference. Top: Two delta pulses with $\xi=6$, spaced to show constructive interference in the production $e^+e^-$ pairs with $p^1=6m$. Bottom: The same, but spaced to show destructive interference.
			(\emph{In the quantum simulations shown below, evolution is performed only between the dark gray lines.})}
		\label{classical_sim}
	\end{figure}
	Fig. \ref{classical_sim} shows the classical real-time Hamiltonian simulation. As expected, the pair-production probability for the $p^1=6m$ mode can experience constructive or destructive interference  depending on the light-front time delay $T$ separating the pulses. The $p^1=0$ and $p^1=12m$ modes experience equal enhancements asymptotically. The various oscillatory behaviors appearing before and after the second pulse are captured by the perturbative formulae. After the first pulse, Eq.~(\ref{lessT}) shows that both the frequency and amplitude of oscillations are sensitive to the combination $\alpha+p^-$, which is quadratic in $p^1$. If $p^1 = m\xi$, then $\alpha+p^-$ is minimized and we see relatively shorter frequencies and larger amplitudes. If $p^1 \neq m\xi$, then we see longer frequency and shorter amplitude oscillations (as in the $p^1=0$ and $p^1=12m$ modes). From Eq.~(\ref{moreT}) we see that the oscillation frequencies after the second pulse depend on the combination $\alpha+\beta+p^-$. Although still quadratic in $p^1$, this expression is minimized and equal to $m^2/p^+$, independent of $\xi$, when $p^1=2m\xi$. This is why the $p^1=12m$ mode has a lower frequency than the $p^1=0$ and $p^1=6m$ modes. The amplitudes of the modes now have a more complicated dependence in $p^1$. Roughly speaking, a factor of $\frac{1}{(\alpha+\beta+p^-)^2}$ leads to large amplitudes for $p^1 \approx 2m\xi$. Away from this minimum, however, the amplitudes are suppressed by ${\cal O}(1/\xi^2)$. Correspondingly for the $p^1=0$ and $p^1=6m$ modes we see much smaller amplitudes: the second pulse effectively ``stops'' the oscillations. For these modes far from $2m\xi$, adiabatic turn-off is therefore  less significant.
	
	The simulations allow us to  visualize interference effects in real time, which are most important in the $p^1=m\xi$ mode. The first pulse gives a kick to the $p^1=m\xi$ mode and its probability starts to oscillate.  The second pulse stabilizes the oscillation of this mode near its maximum. At the same time, the application of two kicks also yields an amplitude in the $p^1=2m\xi$ mode. 
	
\begin{figure}[h]
		\centering\includegraphics[scale=0.42]{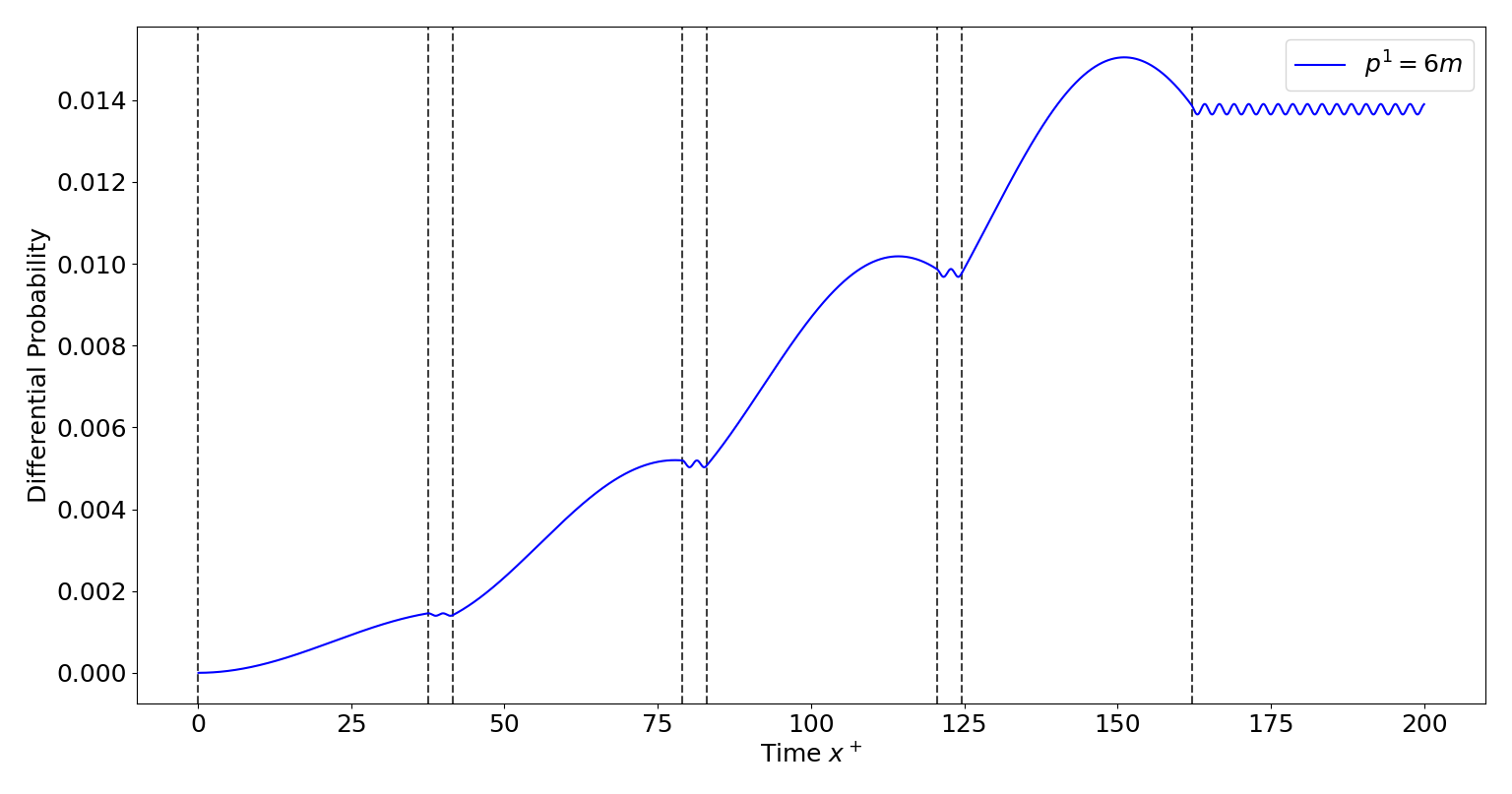}
		\caption{Eight consecutive delta pulses with $\xi=6$ are represented by dark gray vertical lines. The signs of the pulses  appear in the order $++--++--$. A careful selection of pulse spacings can lead to substantial enhancement of the pair production probability in specific momentum modes.}
		\label{mult_enhance}
	\end{figure}

	As a side note, if we introduce a third field with strength $-m\xi$, we might expect to obtain further constructive interference in the $p^1=m\xi$ mode. In other words, if we vary the net field between $m\xi$ with consecutive pulses, we expect to see multiple constructive interference enhancements in the $p^1=m\xi$ mode pair-production probabilities. Fig.~\ref{mult_enhance} exhibits this qualitative behavior \footnote{For the reasons discussed above, we did not include adiabatic turn-on or turn-off in making Fig.~\ref{mult_enhance}, and so we only display the dynamical time between pulses.}  using a background field of the form
	\begin{equation}
		e\mathcal{A}^1(x^+) = m\xi[\theta(x^+) + \theta(x^+ - T_1) - \theta(x^+ - T_2) - \theta(x^+ - T_3) + \theta(x^+ - T_4) + \theta(x^+ - T_5) - \cdots].
	\end{equation}
	The light-front time differences $T_{odd}-T_{even}$ and $T_{even}-T_{odd}$ were kept constant. Subsequent enhancements exhibit somewhat shorter frequency oscillations, so fine-tuning the pulse separations could lead to greater net constructive interference.
	
	\section{Quantum Results}
	\label{sec_quantum_results}
	
	Near-term quantum devices are limited by quantum noise. This means we cannot precisely perform the same classical simulation on a quantum device. The equivalent number of qubits needed for the Breit-Wheeler simulation of Fig. \ref{classical_sim} is $n=14$.\footnote{$n=14$ qubits $=$ (2 photon momenta)$\times$(2 polarizations) + (3 electron momenta)$\times$(2 helicities)  + (3 positron momenta)$\times$(2 helicities).} Based on the discussion below Eq. (\ref{paulis}), we would then expect of order $3(2 \times 2) (2 \times 2) (2 \times 2) = 192$ distinct Pauli strings in the Hamiltonian. For a rough lower bound estimate, a single Trotter step could be implemented with $\gtrsim 600$ gates. Moreover, the majority of Trotter steps would require long-reaching CNOT gates which would introduce many SWAP gates and hence more  noise. For this reason, we will return to the three qubit quantum simulation discussed in Sec. 3, achieved by truncating to fewer helicities and polarizations and momenta.
	
	Beyond gate errors, the probabilities we can actually measure are limited by shot noise. This is a problem particularly for real theories characterized by a weak coupling, like QED, where interesting processes may have quite low probabilities. To avoid requiring an infeasible number of shots, we artificially boost the coupling constant $e=0.303 \to 60$ and decrease the lattice parameter $L \to \frac{6\pi}{m}$. For the electron momentum, we let $p^+ = p^1 = \frac{8m}{3}$. (It still is the case that the positron momentum has $q^+=p^+$ and $q^1=-p^1$. For the photon, $k^+=2p^+$ as before.) With this choices we are still effectively near the continuum limit, since the momentum lattice spacing $\frac{2\pi}{L} = \frac{m}{3}$ is an eighth of the typical momentum magnitude.
	
\begin{figure}[h]
		\centering\includegraphics[scale=0.666]{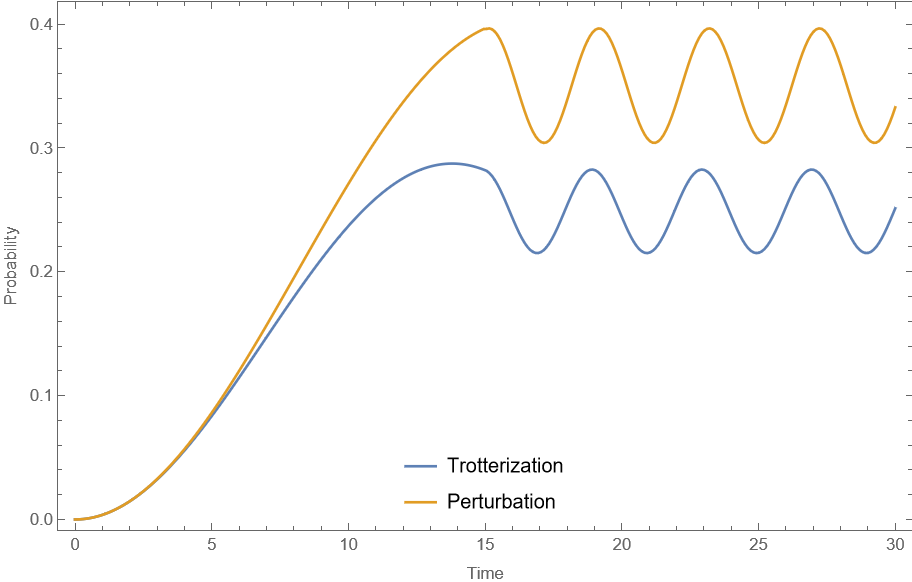}
		\caption{Comparison of real-time pair-production probabilities computed with perturbation theory and Trotterization for artificially large coupling. An initial delta pulse is encountered at time zero, and then the probability oscillation amplitudes change when the second pulse is encountered. At $e=60$, perturbation theory still yields probabilities below one, and the two methods agree qualitatively. Perturbation theory  provides a useful understanding of the process, despite the large coupling, because the phase space volume is small (both for the final state and the Hilbert space truncation as a whole, which prevents higher order effects from becoming large.)}
		\label{pert_vs_trot}
	\end{figure}

	Although taking $e=60$ appears decidedly non-perturbative, what our simulations compute are  exclusive probabilities in small cells of Fock space, or effectively differential probabilities times small phase space factors. By retaining only a few states in the full Fock space, each with a small volume $(\frac{2\pi}{L})^3$, the truncated quantum mechanical model is still perturbative despite the large value of $e$. Tree-level results in the full QFT may then be obtained by rescaling. We emphasize however that this approach will break down if many more Fock states are kept in the simulation (increasing the amplitudes of second-order and higher transitions in the truncated model.) Expanding the truncated theory in this way is of course essential to match with continuum SFQED beyond the leading order and to see more interesting dynamics, so care will have to be taken to design simulations that can operate at the physical coupling without requiring prohibitively large shot counts. In any event, for our purposes an upscaling of the coupling is sufficient. Fig. \ref{pert_vs_trot} shows that for $e=60$ we still obtain qualitatively reasonable agreement between Trotterization and first-order perturbation theory in the truncated quantum mechanics (or the tree level prediction of full SFQED) in this case.
	
	The single Trotter step circuit in Fig. \ref{circuit}, once expressed in native IBM gates, is 40 gates long. While the classical simulation discussed previously was performed over 21000 timesteps, most of which were adiabatic turn-on/off steps, the quantum simulation must be limited to a few hundred gates total in order to avoid total decoherence. To this end, we omit adiabatic turn-on/off in the quantum simulations. Instead, we fix the initial state to the free photon state $\ket{001}$, which as we have discussed previously is a good approximation for the physics of interest. We also begin the time evolution at the (light-front) time of collision between the photon and the first pulse and end it just after the collision with the second pulse (e.g. the quantum simulation is performed only between the dark gray lines of Fig.~\ref{classical_sim}). Within this region, the pair-production probability oscillates. Due to NISQ limitations, we can only afford to capture a small portion of these oscillations, such as half a period. Using 10 Trotter steps, we show how the probability changes from minimum to maximum. To fully capture the noisy dynamics, we also record the $\gamma \to \gamma$ probabilities.
	
	\begin{figure}
		\centering
		\begin{subfigure}{\linewidth}
			\centering
			\includegraphics[scale=0.42]{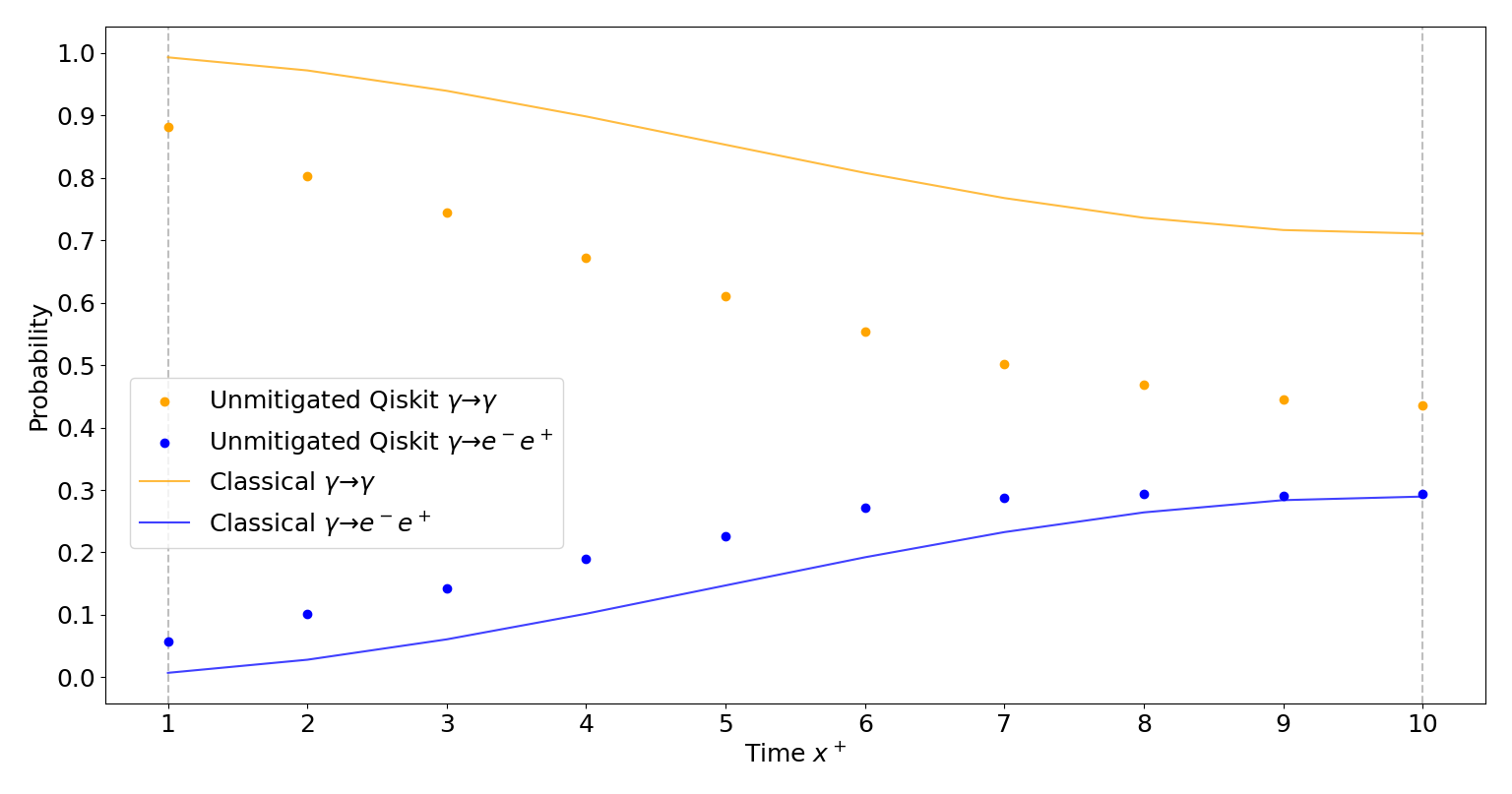}  
			\caption{Results from Qiskit's noisy simulator reflect modest errors, more significant in $\gamma \to \gamma$  than $\gamma \to e^- + e^+$ probabilities.}
			\label{qiskit_unmitigated}
		\end{subfigure}
		
		\begin{subfigure}{\linewidth}
			\centering
			\includegraphics[scale=0.42]{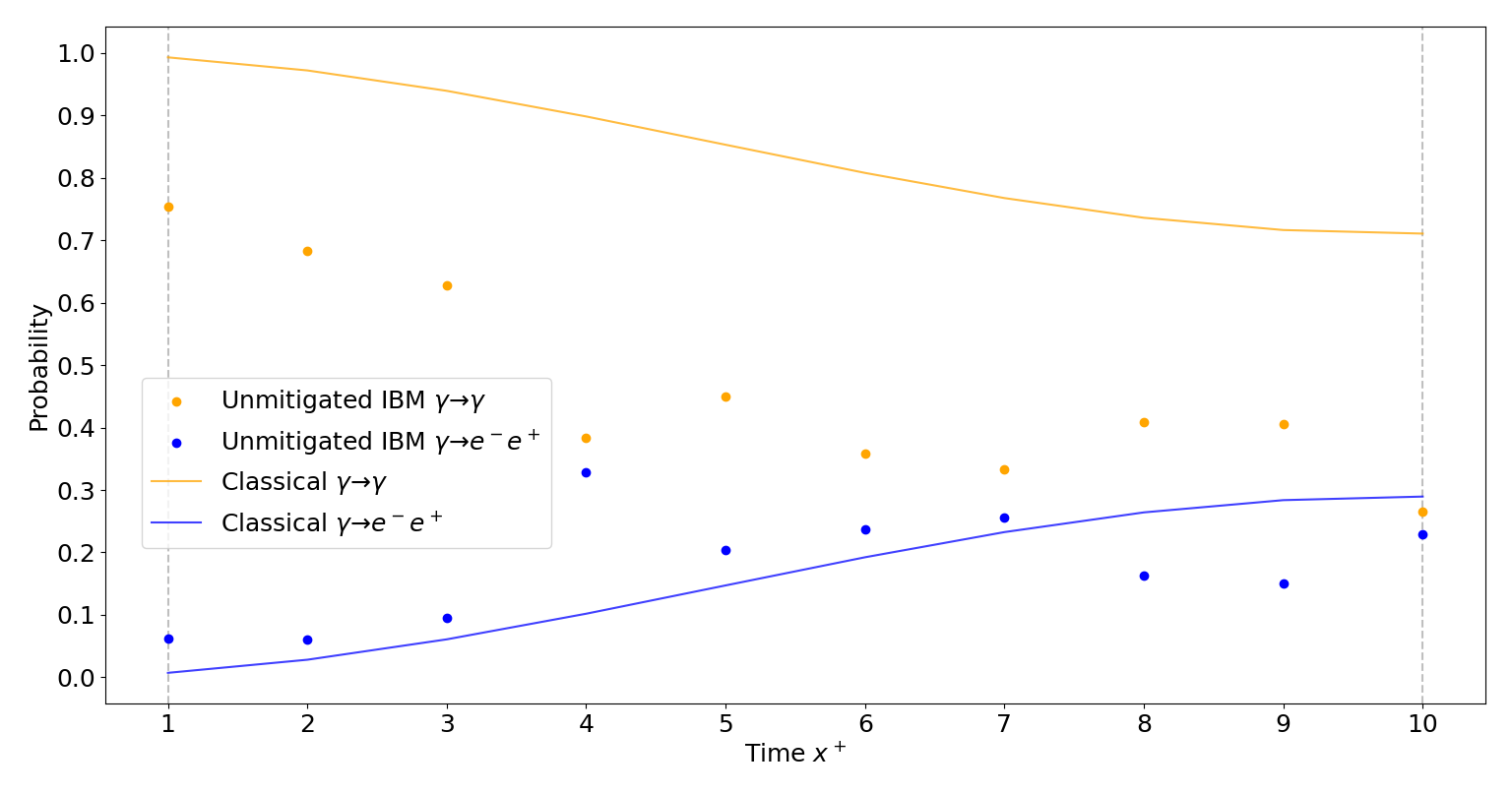}  
			\caption{Data from IBM's quantum hardware reflects strong  noise effects, particularly in $\gamma \to \gamma$ probabilities.}
			\label{ibm_unmitigated}
		\end{subfigure}
		\caption{Unmitigated quantum results for nonlinear Breit-Wheeler pair-production. Pulses with $m\xi = \frac{8m}{3}$ are encountered at $x^+=1$ and $x^+=10$. The coupling is held fixed at $e=60$. The electron transverse momentum mode is $p_1=m\xi$.}
		\label{unmitigated_sim}
	\end{figure}
	
	It is convenient to estimate the impact of noise first on a classical simulator, using a noise model based on historical measurements. Results using Qiskit's \verb|FakeNairobiV2| noisy simulator \cite{Qiskit} are shown in Fig. \ref{qiskit_unmitigated}. The data suggest that we should expect at least 10\% errors already in the initial timestep, increasing with time.
	
	Fig. \ref{ibm_unmitigated} shows raw quantum data from \verb|ibm_nairobi| \cite{ibmquantum}. The real noise level we observe is  higher than seen in the noisy  \verb|FakeNairobiV2| simulator. At $x^+=4$ the error in the $\gamma \to \gamma$ probability is already 60\%. Such results are not uncommon from NISQ devices and can be substantially cleaned by mitigation algorithms.
	
	\section{Error Mitigation}
	\label{sec_error_mitigation}
	
	Mitigating against specific sources of device noise is critical to extract useful information from near-term quantum simulations. We employ measurement mitigation and Pauli-twirling \cite{PhysRevA.94.052325}, as well as a modified version of  self-mitigation \cite{rahman2022real, PhysRevLett.127.270502, PhysRevD.107.054512} (mitigation of depolarization noise). The combination of these three error mitigation techniques significantly improved the quantum data as shown in Fig. \ref{ibm_mitigated}. For completeness we discuss each mitigation algorithm in some detail.
	
	\subsubsection*{Measurement Mitigation}
	
	With measurement mitigation we account for readout errors of bitstrings. For example, there is the probability that upon readout, the measured bitstring is the computational basis state $\ket{000}$ as opposed to $\ket{001}$. For $n$ qubits, the mitigation strategy is to run $2^n$ circuits that prepare individual computational basis states and derive statistics for how many counts there are for the $2^n$ possible results.
	
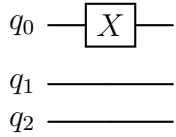
\begin{figure}[h]
		\centering
		\begin{quantikz}
			\lstick{$q_0$} & \gate{X} & \\
			\lstick{$q_1$} & & \\
			\lstick{$q_2$} & &
		\end{quantikz}
		\caption{The measurement mitigation circuit $C_1$ for preparing $\ket{\psi_1}=\ket{001}$ when $n=3$ qubits. The initial state is $\ket{\psi_0}=\ket{000}$ so $C_1\ket{000} = \ket{001}$.}
		\label{meas_mit_ex}
	\end{figure}

	Let $\{\ket{\psi_i}\}$ for $i=0,1,\dots,2^n-1$ be the computational basis. Prepare circuits $C_0,C_1,\dots,C_{2^n-1}$ that prepare each computational basis state as in Fig. \ref{meas_mit_ex}.
	For each circuit $C_j$, measure $|\bra{\psi_i}C_j\ket{\psi_0}|^2$. These probabilities are entries $\mathcal{C}_{i,j} = |\bra{\psi_i}C_j\ket{\psi_0}|^2$ of the calibration matrix $\mathcal{C}$. That is, the measurement probabilities from $C_j$ form the $j$th column of $\mathcal{C}$.
	
	For a general quantum circuit, let $\vec{c}^{true}$ be the column vector of noiseless measurement counts of each basis state. The product $\mathcal{C} \vec{c}^{true}$ distributes the true measurement counts among other basis states such that the calibration matrix models the effects of readout noise. Thus, define $\vec{c}^{meas} = \mathcal{C} \vec{c}^{true}$.
	
	In reality, noisy quantum measurements yield $\vec{c}^{meas}$, so in principle we can obtain the mitigated counts by matrix inversion: $\vec{c}^{true} = \mathcal{C}^{-1} \vec{c}^{meas}$. However, simply inverting $\mathcal{C}$ can lead to negative counts and counts that do not sum to the number of shots $N$. Therefore, we utilize a least-squares minimization protocol. This method takes advantage of the notion that $\vec{c}^{meas} - \mathcal{C}\vec{c}^{true} = 0$. Let $\vec{x}$ be an estimate for $\vec{c}^{true}$. Then the problem becomes finding
	\begin{equation}
		\min_{\vec{x}} \left[ (\vec{c}^{meas}-\mathcal{C}\vec{x})^T (\vec{c}^{meas}-\mathcal{C}\vec{x}) \right]
	\end{equation}
	subject to the constraints that $0 \leq x_i \leq N$ and $|\vec{x}| = N$. The minimization will yield the best estimate of $\vec{c}^{true}$ as predicted by $\mathcal{C}$. The mitigated probabilities are then $\frac{1}{N}\vec{c}^{true}$.
	
	\subsubsection*{Pauli Twirling}
	
\begin{figure}[h]
		\begin{center}
			\begin{quantikz}
				& \ctrl{1} & \\
				& \targ{} &
			\end{quantikz}
			$ = $ \begin{quantikz}
				& \gate{X} & \ctrl{1} & \gate{Y} & \\
				& \gate{Y} & \targ{} & \gate{Z} &
			\end{quantikz}
			$ = $ \begin{quantikz}
				& \gate{X} & \ctrl{1} & \gate{Y} & \\
				& \gate{Z} & \targ{} & \gate{Y} &
			\end{quantikz}
			$ = $ \begin{quantikz}
				& \gate{Z} & \ctrl{1} & \gate{Z} & \\
				& \gate{X} & \targ{} & \gate{X} &
			\end{quantikz}
		\end{center}
		\caption{Sample Pauli twirls of the CNOT gate. For error mitigation, each circuit was randomly compiled with every CNOT twirled using these equivalent  gates. }
		\label{twirls}
	\end{figure}
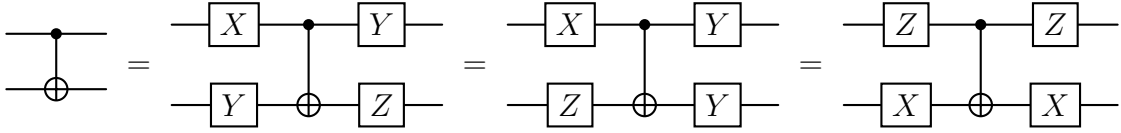

	CNOT gates may have different errors depending on the state that is fed to them. Pauli twirling implements gate identities around CNOTs in order to feed them different states, thus ``symmetrizing'' the noise. Fig. \ref{twirls} shows examples of CNOT identities. In total there are 16 ways to twirl a CNOT gate with extra Pauli gates. We randomly twirl all CNOTs with different identities in each run, then average over the twirled results in post-processing.
	
	\subsubsection*{Depolarization Mitigation}
	The quantum channel often used to  describe depolarizing noise \cite{nielsen00} is
	\begin{equation}
		\tilde{\rho} = p\frac{I}{2^n} + (1-p)\rho.
	\end{equation}
	Under this channel a state $\rho$ may turn into the fully decohered state $\frac{I}{2^n}$ with probability $p$, and with probability $1-p$ the state  remains unchanged. This type of depolarizing noise model is called symmetric depolarization and is the basis for many implementations of ``self-mitigation'' \cite{rahman2022real, PhysRevLett.127.270502, PhysRevD.107.054512}. A more general noise model could take $\rho$ not just to $\frac{I}{2^n}$, but to other general mixed states, which is what we will now consider.
	
	We generalize the depolarizing noise model by introducing a transfer matrix $M$:
	\begin{equation}
		\sum_{j=0}^{2^n-1} M_{ij} P_j^{true} \equiv P^{meas}_i,
		\label{noise_model}
	\end{equation}
	where $P_i^{true}$ ($P^{meas}_i$) is the true (noisy measurement) probability of computational basis state $\ket{\psi_i}$. This implies the following quantum channel:
	\begin{equation}
		\tilde{\rho} = \sum_{i=0}^{2^n-1} \sum_{j=0}^{2^n-1} M_{ij} \rho_{jj} \ket{\psi_i}\bra{\psi_i}.
	\end{equation}
	
\begin{figure}[h]
		\centering
		\begin{quantikz}
			\lstick{$q_0$} & \gate{X} & \gate[3]{I_{noisy}} & \\
			\lstick{$q_1$} & & & \\
			\lstick{$q_2$} & & &
		\end{quantikz} = \begin{quantikz}
			\lstick{$q_0$} & \gate{X} & \gate[3]{e^{-\frac{i}{2}H(x_1^+)\Delta x^+} e^{-\frac{i}{2}H(x_2^+)\Delta x^+} e^{\frac{i}{2}H(x_2^+)\Delta x^+} e^{\frac{i}{2}H(x_1^+)\Delta x^+}} & \\
			\lstick{$q_1$} & & & \\
			\lstick{$q_2$} & & &
		\end{quantikz}
		\caption{The depolarizing mitigation circuit $C_1(x^+_2)$ for the second time step, preparing $\ket{\psi_1}=\ket{001}$ when $n=3$ qubits. The noisy identity is the time evolution circuit used for the quantum simulation followed by its Hermitian conjugate.}
		\label{dep_mit_ex}
	\end{figure}
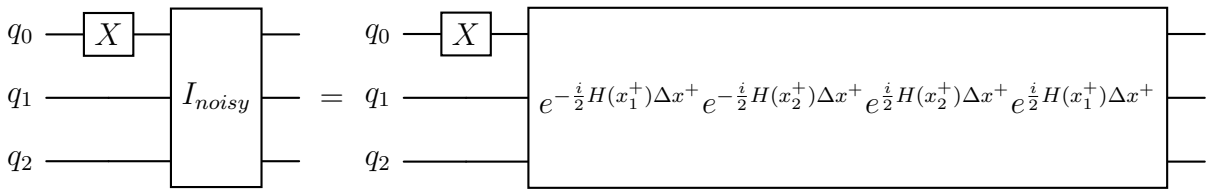

	Symmetric depolarizing noise is the case where $M$ is a one-parameter matrix with entries $M_{ij} = \delta_{ij}(1-2^n\epsilon) + \epsilon$.  $\epsilon$ may be estimated by preparing circuits with similar noise elements as the circuit of interest. Likewise, to estimate the elements of $M_{ij}$ in our model, we construct circuits $C_0,C_1,\dots,C_{2^n-1}$, similar to those for measurement mitigation. However, in addition to preparing each computational basis state, the circuit applies an identity operator with the same gates and the physics circuit, as shown in Fig. \ref{dep_mit_ex}.
	Each noisy identity is the current time evolution operator followed by its inverse. This means that the mitigation circuits $C_i$ are time-dependent $C_i(x^+)$, allowing us to mitigate against time-dependent noise. The transfer matrix elements are then estimated as $M_{ij}(x^+) = |\bra{\psi_i} C_j(x^+) \ket{\psi_0}|^2$.
	
	We  emphasize that in both the symmetric case and our case, Pauli-twirling is essential to convert more general types of noise from the CNOTs into depolarization noise. Thus,  depolarization mitigation is the last step of the error mitigation pipeline. Both the physics and mitigation circuits are twirled and the results are averaged before depolarization mitigation.
	
	Furthermore, since the inserted identity operators are effectively  ``twice the physics circuit," each mitigation circuit is practically a measure for the noise ``squared''. This implies that we should be taking $\sqrt{M}$ as the calibration matrix.\footnote{This is also why the measurement mitigation and depolarization mitigation cannot be done at the same time in our problem, despite both being formulated in terms of a calibration matrix.} That is,
	\begin{equation}
		\vec{c}^{meas} = \sqrt{M} \vec{c}^{true}.
	\end{equation}
	As before, to determine  $\vec{c}^{true}$, we solve an  optimization problem 
	\begin{equation}
		\min_{\vec{x}} \left[ (\vec{c}^{meas}-\sqrt{M}\vec{x})^T W (\vec{c}^{meas}-\sqrt{M}\vec{x}) \right]
	\end{equation}
	subject to the same constraints on $\vec{x}$ as before. The weights $W$ are taken to be $W=\diag(\frac{1}{p_i^2})$, where $p_i = |\bra{\psi_i} C_i \ket{\psi_0}|^2$.
	
	We checked that symmetric depolarization noise mitigation following Pauli twirling did not perform well on our problem: the noise profile indicated some states ``thermalize" with each other, but not all together. On the other hand, the construction of a complete calibration matrix for depolarization mitigation is computationally expensive and does not scale well with problem size. For the small size studied here it was not critical to optimize it further, but it would be interesting to explore the performance restricted classes of calibration transformations constructed from fewer parameters and mitigation circuit measurements.
	
	\begin{figure}
		\centering
		\begin{subfigure}{\linewidth}
			\centering
			\includegraphics[scale=0.42]{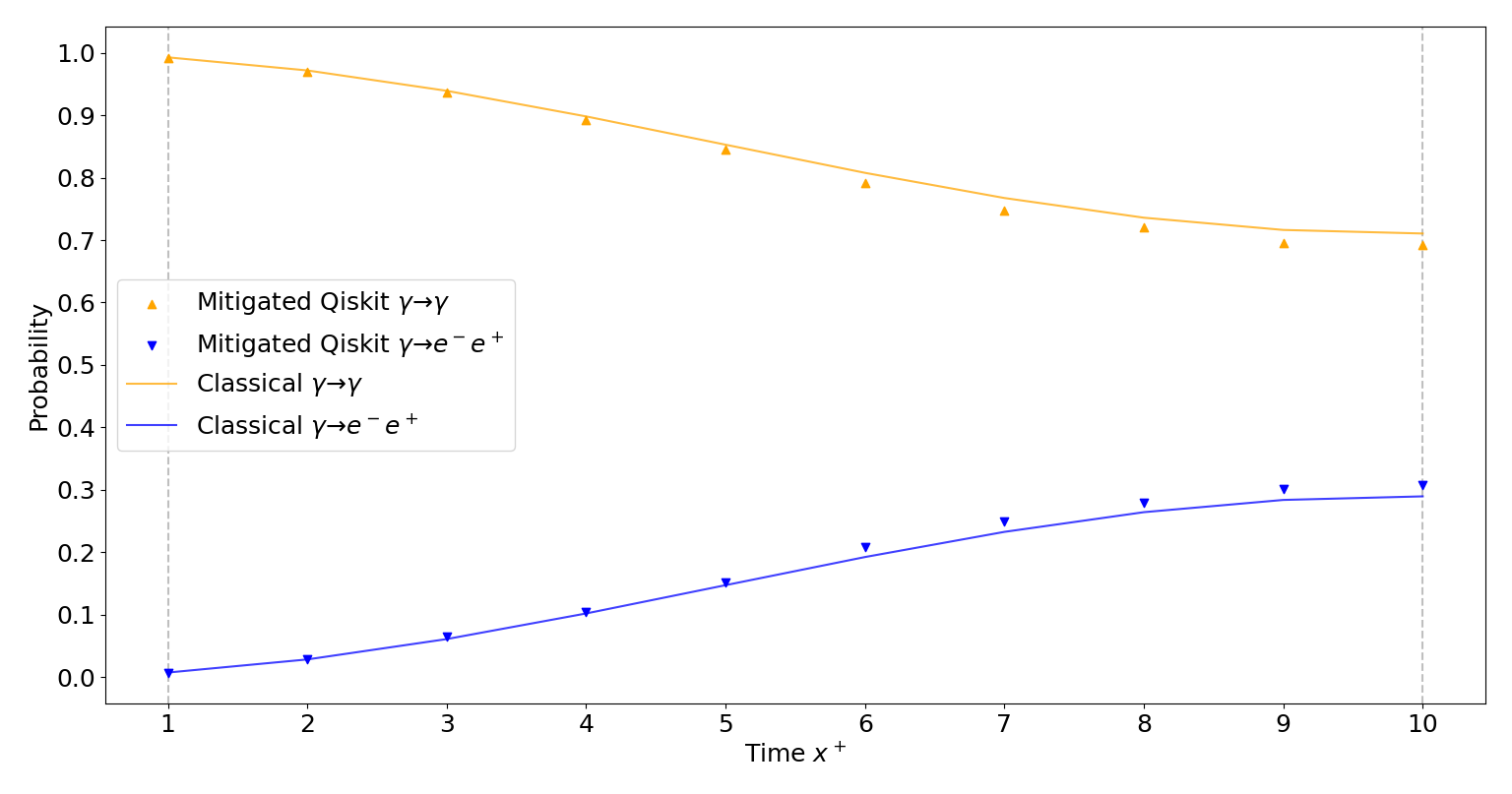}  
			\caption{Mitigated data from Qiskit's fake simulator shows excellent agreement with classical simulation results.}
			\label{qiskit_mitigated}
		\end{subfigure}
		
		\begin{subfigure}{\linewidth}
			\centering
			\includegraphics[scale=0.42]{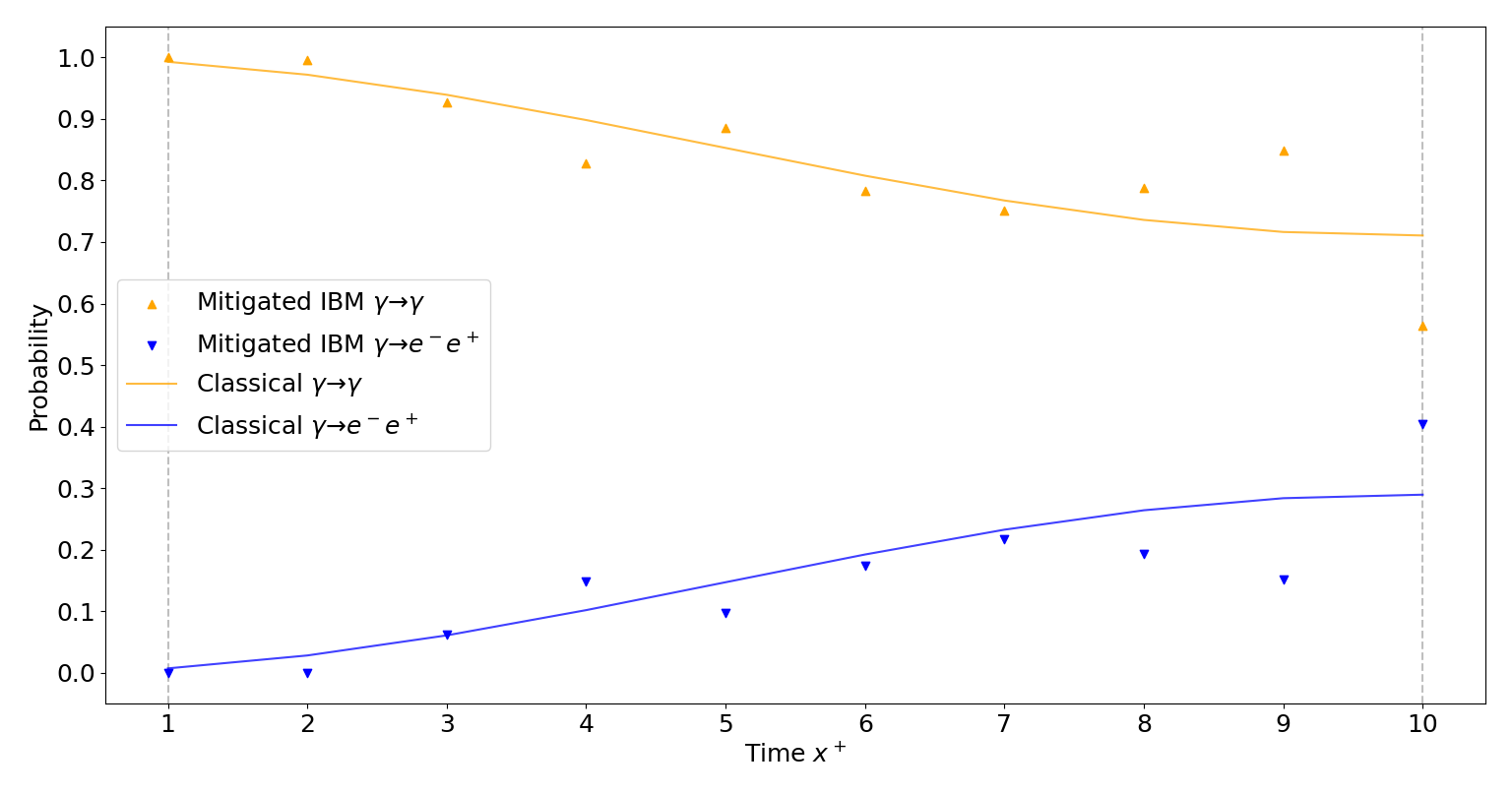}  
			\caption{Mitigated data from IBM's quantum hardware agrees well with classical results until $x^+=9$, suppressing high levels of errors seen in raw data.}
			\label{ibm_mitigated}
		\end{subfigure}
		\caption{Mitigated quantum results for nonlinear Breit-Wheeler pair-production. Pulses with $m\xi = \frac{8m}{3}$ are encountered at $x^+=1$ and $x^+=10$. Robust measurement and depolarization mitigation techniques suppress errors at early time steps, but become less effective at late time steps.}
		\label{mitigated_sim}
	\end{figure}
	
	\subsubsection*{Mitigated Results}
	
	For each time step, we performed 10 Pauli-twirls on the quantum simulation circuit as well as each of the $2^3=8$ mitigation circuits. Each time step was sent as a different set of circuits to \verb|ibm_nairobi| at different times, so in addition we ran 8 measurement mitigation circuits each time step. Therefore, 98 circuits were sent to \verb|ibm_nairobi| at each time step.

	As before we can compare with Qiskit's noisy \verb|FakeNairobiV2| simulator. Fig. \ref{qiskit_mitigated} shows the mitigated noisy simulator results. The mitigation is substantial and suggests that the real quantum simulation should also see improved results vs the raw data.
	
	Fig. \ref{ibm_mitigated} shows mitigated quantum data. As expected, the error mitigation protocol greatly improves the simulation results. The largest percent error we see is about 15\% of the classical simulation. However, substantial errors do not occur until the last two time steps. This suggests that had we performed a longer or finer time evolution, decoherence may have been too strong to be well-mitigated.
	
	\section{Conclusion}
	\label{sec_conclusion}
	
	Quantum field theory  in the presence of strong background fields is a rich arena where quantum computers may provide valuable new access. We have seen that for a simple strong-field QED process -- nonlinear Breit-Wheeler pair-production in a sequence of laser pulses -- the noisy quantum computers of today are already able to simulate the dynamics reliably with the help of error mitigation routines. Due to the polynomial scaling in circuit depth,  the simulation of SFQED with many particles and strong backreaction presents an interesting, plausible long-term goal. 
	
	Capturing more complex electromagnetic cascade processes requires retaining a much larger Hilbert space. With a less drastic truncation one could also study helicity and polarization effects in cascades, as well as begin to tackle the regime of ultra-strong fields, where few other tools exist. The light-front Fock space formulation used in this work can be applied directly to more complicated processes with bigger Hilbert spaces, and the operator-circuit map can be straightforwardly extended.

	It is also possible to simulate O(30-50) qubit quantum computers with state-of-the-art GPUs and supercomputers~\cite{Bayraktar:2023sjt}. Larger-scale SFQED quantum circuits could be designed on these simulators with adjustable noise. This capability could be useful to  develop efficient circuits for future quantum computers, and to establish noise targets for interesting physics processes. Investigations in this direction are ongoing.

	\section{Acknowledgements}
	We thank Bryce Gadway for discussions and suggesting the possibility of greater enhancements from many pulses. This work was supported in part by the U.S. Department of Energy, Office of Science, Office of High Energy 
	Physics under award number DE-SC0015655 and by its
	QuantISED program under an award for the Fermilab
	Theory Consortium ``Intersections of QIS and Theoretical Particle Physics." We acknowledge the use of IBM
	Quantum services for this work. The views expressed
	are those of the authors, and do not reflect the official
	policy or position of IBM or the IBM Quantum team.

	\nocite{*}
	\bibliographystyle{utphys}
	\bibliography{bib}
	
	\appendix
	
	\section{Quantized Hamiltonian}
	\label{app_A}
	
	The SFQED Hamiltonian density was derived in Sec.~\ref{sec_sfqed_hamiltonian}. Integrating the density over light-front space yields the Hamiltonian. We also expand the fields in momentum modes. The fields are given in Eq. (\ref{fields_exp}) and the commutation relations are given in Eq. (\ref{commutators}). In this appendix, we derive a few  sample terms in the mode expansion of the Hamiltonian.
	
	\subsection{Fermion Energy Hamiltonian}
	
	We begin with the free fermion energy $H_\psi$. We will need to compute expressions of the form $\frac{f(x^-)}{\partial_-}$. In momentum space the Green function contributes factors of $\frac{1}{ip^+}$. We have
	\begin{equation}
		\frac{\psi_-}{\partial_-} = \int \frac{dp^+ dp^\perp}{(2\pi)^3} \sum_{s=\pm\frac{1}{2}} \frac{2e^{-i\mathsf{p}\mathsf{x}}}{-ip^+} w^s a_p^s + \frac{2e^{i\mathsf{p}\mathsf{x}}}{ip^+} w^{-s} b_p^{s\dag}
	\end{equation}
	and
	\begin{equation}
		\frac{\partial_i \partial_i \psi_-}{\partial_-} = \int \frac{dp^+ dp^\perp}{(2\pi)^3} \sum_{s=\pm\frac{1}{2}} \frac{2(ip^\perp)^2e^{-i\mathsf{p}\mathsf{x}}}{-ip^+} w^s a_p^s + \frac{2(-ip^\perp)^2e^{i\mathsf{p}\mathsf{x}}}{ip^+} w^{-s} b_p^{s\dag}.
	\end{equation}
	Using orthogonality of the spinors,
	\begin{equation}
		w^{\pm s\dag} w^{\pm s'} = \delta_{s,s'} \qquad w^{\pm s\dag} w^{\mp s'} = \delta_{-s,s'},
	\end{equation}
	and integrating over space, we find
	\begin{IEEEeqnarray*}{rCl}
		H_\psi & = & m^2 \int \frac{dp^+ dp^\perp}{(2\pi)^3} \int d{p'}^+ d{p'}^\perp \sum_{s=\pm\frac{1}{2}} \\
		& & \hspace{3cm} \times \left( \frac{\delta^3(\mathsf{p}-\mathsf{p}')}{{p'}^+} a_p^{s\dag} a_{p'}^s - \frac{\delta^3(\mathsf{p}-\mathsf{p}')}{{p'}^+} b_p^s b_{p'}^{s\dag} - \frac{\delta^3(\mathsf{p}+\mathsf{p}')}{{p'}^+} a_p^{s\dag} b_{p'}^{-s\dag} + \frac{\delta^3(\mathsf{p}+\mathsf{p}')}{{p'}^+} b_p^s a_{p'}^{-s} \right) \\
		& + & \int \frac{dp^+ dp^\perp}{(2\pi)^3} \int d{p'}^+ d{p'}^\perp \sum_{s=\pm\frac{1}{2}} ({p'}^\perp)^2 \\
		& & \hspace{3cm} \times \left( \frac{\delta^3(\mathsf{p}-\mathsf{p}')}{{p'}^+} a_p^{s\dag} a_{p'}^s - \frac{\delta^3(\mathsf{p}-\mathsf{p}')}{{p'}^+} b_p^s b_{p'}^{s\dag} - \frac{\delta^3(\mathsf{p}+\mathsf{p}')}{{p'}^+} a_p^{s\dag} b_{p'}^{-s\dag} + \frac{\delta^3(\mathsf{p}+\mathsf{p}')}{{p'}^+} b_p^s a_{p'}^{-s} \right) \\
		& & \yesnumber
	\end{IEEEeqnarray*}
	The delta functions enforce momentum conservation, $p^+ = {p'}^+$ and $p^\perp = {p'}^\perp$. However, $\delta^3(\mathsf{p}+\mathsf{p}')$ and $p^+>0$ indicates that integrals with $p^+ \leq 0$ vanish. Only $\delta^3(\mathsf{p}-\mathsf{p}')$ terms contribute:
	\begin{IEEEeqnarray*}{rCl}
		H_\psi & = & m^2 \int \frac{dp^+ dp^\perp}{(2\pi)^3} \sum_{s=\pm\frac{1}{2}} \left( \frac{1}{p^+} a_p^{s\dag} a_p^s - \frac{1}{p^+} b_p^s b_p^{s\dag} \right) + \int \frac{dp^+ dp^\perp}{(2\pi)^3} \sum_{s=\pm\frac{1}{2}} (p^\perp)^2 \left( \frac{1}{p^+} a_p^{s\dag} a_p^s - \frac{1}{p^+} b_p^s b_p^{s\dag} \right) \\
		& = & \int \frac{dp^+ dp^\perp}{(2\pi)^3} \sum_{s=\pm\frac{1}{2}} \left( \frac{(p^\perp)^2 + m^2}{p^+} a_p^{s\dag} a_p^s - \frac{(p^\perp)^2 + m^2}{p^+} b_p^s b_p^{s\dag} \right) \\
		& = & \int \frac{dp^+ dp^\perp}{(2\pi)^3} \sum_{s=\pm\frac{1}{2}} p^- \left( a_p^{s\dag} a_p^s - b_p^s b_p^{s\dag} \right). \yesnumber
	\end{IEEEeqnarray*}
	Finally, we normal order by anticommuting the positron operators in the usual way. The final result for this term is
	\begin{equation}
		:\mathrel{H_\psi}: \ = \int \frac{dp^+ dp^\perp}{(2\pi)^3} \sum_{s=\pm\frac{1}{2}} p^- \left( a_p^{s\dag} a_p^s + b_p^{s\dag} b_p^s \right)
	\end{equation}
	which sums the light-front energies of the fermions.
	
	\subsection{Fermion-Photon Interaction Hamiltonian}
	
	The other terms for which we will derive the explicit mode expansion are the interaction Hamiltonian used in our simulations. These terms come from
	\begin{equation}
		\mathcal{H}_{\psi A} \supset \frac{ime}{2} \psi_-^\dag \gamma^j \left( \frac{A^j\psi_-}{\partial_-} \right) - \frac{ime}{2} \psi_-^\dag \gamma^j A^j \left( \frac{\psi_-}{\partial_-} \right)
	\end{equation}
	(Here repeated indices, although all raised, are understood to be summed over.) 
	It is convenient to define the following coefficients $R$:
	\begin{equation}
		R^{s,j} \equiv w^{s\dag} \gamma^j w^{-s} \qquad R^{+\frac{1}{2},j} = \begin{cases}
			1 & j=1 \\
			i & j=2
		\end{cases} \qquad R^{-\frac{1}{2},j} = \begin{cases}
			-1 & j=1 \\
			i & j=2
		\end{cases}
	\end{equation}
	Integrating over light-front space we obtain
	\begin{IEEEeqnarray*}{rCl}
		\frac{1}{2} \int d^3\mathsf{x} \ \psi_-^\dag \gamma^j \left( \frac{A^j\psi_-}{\partial_-} \right) & = & 2i \int \frac{dp^+ dp^\perp}{(2\pi)^3} \int \frac{dk^+ dk^\perp}{(2\pi)^3\sqrt{k^+}} \int d{p'}^+ d{p'}^\perp \sum_{s=\pm\frac{1}{2}} \sum_{j=1,2} \\
		& \times & \left( \frac{\delta^3(-\mathsf{p}+\mathsf{k}+\mathsf{p}')}{k^+ + {p'}^+} R^{s,j} a_p^{s\dag} c_k^j a_{p'}^{-s} - \frac{\delta^3(\mathsf{p}+\mathsf{k}+\mathsf{p}')}{k^+ + {p'}^+} R^{s,j} a_p^{s\dag} c_k^{j\dag} b_{p'}^{s\dag} \right. \\
		& + & \frac{\delta^3(-\mathsf{p}+\mathsf{k}-\mathsf{p}')}{k^+ - {p'}^+} R^{s,j} a_p^{s\dag} c_k^j b_{p'}^{s\dag} - \frac{\delta^3(\mathsf{p}+\mathsf{k}-\mathsf{p}')}{k^+ - {p'}^+} R^{s,j} a_p^{s\dag} c_k^{j\dag} a_{p'}^{-s} \\
		& + & \frac{\delta^3(\mathsf{p}+\mathsf{k}+\mathsf{p}')}{k^+ + {p'}^+} R^{-s,j} b_p^s c_k^j a_{p'}^s - \frac{\delta^3(-\mathsf{p}+\mathsf{k}+\mathsf{p}')}{k^+ + {p'}^+} R^{-s,j} b_p^s c_k^{j\dag} b_{p'}^{-s\dag} \\
		& + & \left. \frac{\delta^3(\mathsf{p}+\mathsf{k}-\mathsf{p}')}{k^+ - {p'}^+} R^{-s,j} b_p^s c_k^j b_{p'}^{-s\dag} - \frac{\delta^3(-\mathsf{p}+\mathsf{k}-\mathsf{p}')}{k^+ - {p'}^+} R^{-s,j} b_p^s c_k^{j\dag} a_{p'}^s \right) \\
		& & \yesnumber
	\end{IEEEeqnarray*}
	and
	\begin{IEEEeqnarray*}{rCl}
		\frac{1}{2} \int d^3\mathsf{x} \ \psi_-^\dag \gamma^j A^j \left( \frac{\psi_-}{\partial_-} \right) & = & 2i \int \frac{dp^+ dp^\perp}{(2\pi)^3} \int \frac{dk^+ dk^\perp}{(2\pi)^3\sqrt{k^+}} \int d{p'}^+ d{p'}^\perp \sum_{s=\pm\frac{1}{2}} \sum_{j=1,2} \\
		& \times & \left( \frac{\delta^3(-\mathsf{p}+\mathsf{k}+\mathsf{p}')}{{p'}^+} R^{s,j} a_p^{s\dag} c_k^j a_{p'}^{-s} - \frac{\delta^3(\mathsf{p}+\mathsf{k}+\mathsf{p}')}{{p'}^+} R^{s,j} a_p^{s\dag} c_k^{j\dag} b_{p'}^{s\dag} \right. \\
		& - & \frac{\delta^3(-\mathsf{p}+\mathsf{k}-\mathsf{p}')}{{p'}^+} R^{s,j} a_p^{s\dag} c_k^j b_{p'}^{s\dag} + \frac{\delta^3(\mathsf{p}+\mathsf{k}-\mathsf{p}')}{{p'}^+} R^{s,j} a_p^{s\dag} c_k^{j\dag} a_{p'}^{-s} \\
		& + & \frac{\delta^3(\mathsf{p}+\mathsf{k}+\mathsf{p}')}{{p'}^+} R^{-s,j} b_p^s c_k^j a_{p'}^s - \frac{\delta^3(-\mathsf{p}+\mathsf{k}+\mathsf{p}')}{{p'}^+} R^{-s,j} b_p^s c_k^{j\dag} b_{p'}^{-s\dag} \\
		& - & \left. \frac{\delta^3(\mathsf{p}+\mathsf{k}-\mathsf{p}')}{{p'}^+} R^{-s,j} b_p^s c_k^j b_{p'}^{-s\dag} + \frac{\delta^3(-\mathsf{p}+\mathsf{k}-\mathsf{p}')}{{p'}^+} R^{-s,j} b_p^s c_k^{j\dag} a_{p'}^s \right) \\
		& & \yesnumber
	\end{IEEEeqnarray*}
	Now let us integrate $\mathsf{p}'$ while keeping explicit the momentum conservation in each integrand. Once again integrals with $\delta^3(\mathsf{p}+\mathsf{k}+\mathsf{p}')$  vanish since they require a longitudinal momentum component to be nonpositive. We obtain
	\begin{IEEEeqnarray*}{rCl}
		\frac{1}{2} \int d^3\mathsf{x} \ \psi_-^\dag \gamma^j \left( \frac{A^j\psi_-}{\partial_-} \right) & = & 2i \int \frac{dp^+ dp^\perp}{(2\pi)^3} \int \frac{dk^+ dk^\perp}{(2\pi)^3\sqrt{k^+}} \sum_{s=\pm\frac{1}{2}} \sum_{j=1,2} \\
		& \times & \left( \frac{1}{k^+ + {p'}^+} R^{s,j} a_p^{s\dag} c_k^j a_{p'}^{-s} \bigg|_{\mathsf{p}'=\mathsf{p}-\mathsf{k}} + \frac{1}{k^+ - {p'}^+} R^{s,j} a_p^{s\dag} c_k^j b_{p'}^{s\dag} \bigg|_{\mathsf{p}'=\mathsf{k}-\mathsf{p}} \right. \\
		& - & \frac{1}{k^+ - {p'}^+} R^{s,j} a_p^{s\dag} c_k^{j\dag} a_{p'}^{-s} \bigg|_{\mathsf{p}'=\mathsf{p}+\mathsf{k}} - \frac{1}{k^+ + {p'}^+} R^{-s,j} b_p^s c_k^{j\dag} b_{p'}^{-s\dag} \bigg|_{\mathsf{p}'=\mathsf{p}-\mathsf{k}} \\
		& + & \left. \frac{1}{k^+ - {p'}^+} R^{-s,j} b_p^s c_k^j b_{p'}^{-s\dag} \bigg|_{\mathsf{p}'=\mathsf{p}+\mathsf{k}} - \frac{1}{k^+ - {p'}^+} R^{-s,j} b_p^s c_k^{j\dag} a_{p'}^s \bigg|_{\mathsf{p}'=\mathsf{k}-\mathsf{p}} \right) \\
		& & \yesnumber
		\label{derived_cosmetic}
	\end{IEEEeqnarray*}
	and
	\begin{IEEEeqnarray*}{rCl}
		\frac{1}{2} \int d^3\mathsf{x} \ \psi_-^\dag \gamma^j A^j \left( \frac{\psi_-}{\partial_-} \right) & = & 2i \int \frac{dp^+ dp^\perp}{(2\pi)^3} \int \frac{dk^+ dk^\perp}{(2\pi)^3\sqrt{k^+}} \sum_{s=\pm\frac{1}{2}} \sum_{j=1,2} \\
		& \times & \left( \frac{1}{{p'}^+} R^{s,j} a_p^{s\dag} c_k^j a_{p'}^{-s} \bigg|_{\mathsf{p}'=\mathsf{p}-\mathsf{k}} - \frac{1}{{p'}^+} R^{s,j} a_p^{s\dag} c_k^j b_{p'}^{s\dag} \bigg|_{\mathsf{p}'=\mathsf{k}-\mathsf{p}} \right. \\
		& + & \frac{1}{{p'}^+} R^{s,j} a_p^{s\dag} c_k^{j\dag} a_{p'}^{-s} \bigg|_{\mathsf{p}'=\mathsf{p}+\mathsf{k}} - \frac{1}{{p'}^+} R^{-s,j} b_p^s c_k^{j\dag} b_{p'}^{-s\dag} \bigg|_{\mathsf{p}'=\mathsf{p}-\mathsf{k}} \\
		& - & \left. \frac{1}{{p'}^+} R^{-s,j} b_p^s c_k^j b_{p'}^{-s\dag} \bigg|_{\mathsf{p}'=\mathsf{p}+\mathsf{k}} + \frac{1}{{p'}^+} R^{-s,j} b_p^s c_k^{j\dag} a_{p'}^s \bigg|_{\mathsf{p}'=\mathsf{k}-\mathsf{p}} \right). \\
		& & \yesnumber
	\end{IEEEeqnarray*}
	
	As a cosmetic change, we write Eq.~(\ref{derived_cosmetic}) in terms of $\frac{1}{p^+}$.  Combining terms we obtain
	\begin{IEEEeqnarray*}{rCl}
		H_{\psi A} & \supset & \int \frac{dp^+ dp^\perp}{(2\pi)^3} \int \frac{dk^+ dk^\perp}{(2\pi)^3\sqrt{k^+}} \sum_{s=\pm\frac{1}{2}} \sum_{j=1,2} em \\
		& \times & \left( \left( \frac{1}{{p'}^+} - \frac{1}{p^+} \right) R^{s,j} a_p^{s\dag} c_k^j a_{p'}^{-s} \bigg|_{\mathsf{p}'=\mathsf{p}-\mathsf{k}} - \left( \frac{1}{{p'}^+} + \frac{1}{p^+} \right) R^{s,j} a_p^{s\dag} c_k^j b_{p'}^{s\dag} \bigg|_{\mathsf{p}'=\mathsf{k}-\mathsf{p}} \right. \\
		& + & \left( \frac{1}{{p'}^+} - \frac{1}{p^+} \right) R^{s,j} a_p^{s\dag} c_k^{j\dag} a_{p'}^{-s} \bigg|_{\mathsf{p}'=\mathsf{p}+\mathsf{k}} - \left( \frac{1}{{p'}^+} - \frac{1}{p^+} \right) R^{-s,j} b_p^s c_k^{j\dag} b_{p'}^{-s\dag} \bigg|_{\mathsf{p}'=\mathsf{p}-\mathsf{k}} \\
		& - & \left. \left( \frac{1}{{p'}^+} - \frac{1}{p^+} \right) R^{-s,j} b_p^s c_k^j b_{p'}^{-s\dag} \bigg|_{\mathsf{p}'=\mathsf{p}+\mathsf{k}} + \left( \frac{1}{{p'}^+} + \frac{1}{p^+} \right) R^{-s,j} b_p^s c_k^{j\dag} a_{p'}^s \bigg|_{\mathsf{p}'=\mathsf{k}-\mathsf{p}} \right) \yesnumber
	\end{IEEEeqnarray*}
	
	Once again we normal order. As an arbitrary convention we write the photon creation (annihilation) operators in the leftmost (rightmost) position. 
	In this case, the delta functions in the normal ordering process produces terms that vanish. Rearranging  terms  to place  Hermitian conjugates next to each other, the contribution to the normal-ordered Schr\"{o}dinger-picture interaction Hamiltonian reads
	\begin{IEEEeqnarray*}{rCl}
		:\mathrel{H_{\psi A}}: & \supset & \int \frac{dp^+ dp^\perp}{(2\pi)^3} \int \frac{dk^+ dk^\perp}{(2\pi)^3\sqrt{k^+}} \sum_{s=\pm\frac{1}{2}} \sum_{j=1,2} em \\
		& \times & \left( \left( \frac{1}{{p'}^+} - \frac{1}{p^+} \right) R^{s,j} c_k^{j\dag} a_p^{s\dag} a_{p'}^{-s} \bigg|_{\mathsf{p}'=\mathsf{p}+\mathsf{k}} + \left( \frac{1}{{p'}^+} - \frac{1}{p^+} \right) R^{s,j} a_p^{s\dag} a_{p'}^{-s} c_k^j \bigg|_{\mathsf{p}'=\mathsf{p}-\mathsf{k}} \right. \\
		& + & \left( \frac{1}{{p'}^+} - \frac{1}{p^+} \right) R^{-s,j} c_k^{j\dag} b_{p'}^{-s\dag} b_p^s \bigg|_{\mathsf{p}'=\mathsf{p}-\mathsf{k}} + \left( \frac{1}{{p'}^+} - \frac{1}{p^+} \right) R^{-s,j} b_{p'}^{-s\dag} b_p^s c_k^j \bigg|_{\mathsf{p}'=\mathsf{p}+\mathsf{k}} \\
		& + & \left. \left( \frac{1}{{p'}^+} + \frac{1}{p^+} \right) R^{-s,j} c_k^{j\dag} b_p^s a_{p'}^s \bigg|_{\mathsf{p}'=\mathsf{k}-\mathsf{p}} - \left( \frac{1}{{p'}^+} + \frac{1}{p^+} \right) R^{s,j} a_p^{s\dag} b_{p'}^{s\dag} c_k^j \bigg|_{\mathsf{p}'=\mathsf{k}-\mathsf{p}} \right). \yesnumber
	\end{IEEEeqnarray*}
	
	To pass to the interaction picture, we  evolve the creation and annihilation operators with $U_0$ (see Eq.~(\ref{schro_time_evo0})):
	\begin{IEEEeqnarray*}{rCl}
		U_0^\dag a^s_p U_0 = e^{-\frac{ip^-x^+}{2}} e^{-if(p)} a^s_p & \qquad & U_0^\dag a^{s\dag}_p U_0 = e^{\frac{ip^-x^+}{2}} e^{if(p)} a^{s\dag}_p \\
		U_0^\dag b^s_p U_0 = e^{-\frac{ip^-x^+}{2}} e^{-ig(p)} b^s_p & \qquad & U_0^\dag b^{s\dag}_p U_0 = e^{\frac{ip^-x^+}{2}} e^{ig(p)} b^{s\dag}_p \\
		U_0^\dag c^j_k U_0 = e^{\frac{-ik^-x^+}{2}} c^j_k & \qquad & U_0^\dag c^{j\dag}_k U_0 = e^{\frac{ik^-x^+}{2}} c^{j\dag}_k. \yesnumber
	\end{IEEEeqnarray*}
	It turns out that $f(p)$ and $g(p)$ are the Volkov phases
	\begin{equation}
		f(p) = \int_{0}^{x^+} \frac{dy^+}{2}\left[ e\mathcal{A}^- - \frac{2ep^i}{p^+}\mathcal{A}^i + \frac{e^2 \mathcal{A}^i \mathcal{A}^i}{p^+} \right] \qquad g(p) = \int_{0}^{x^+} \frac{dy^+}{2}\left[ -e\mathcal{A}^- + \frac{2ep^i}{p^+}\mathcal{A}^i + \frac{e^2 \mathcal{A}^i \mathcal{A}^i}{p^+} \right].
	\end{equation}
	These are part of the solutions to the Dirac equation in the presence of a background vector field $\mathcal{A}_\mu$ (specifically a null field, for which ${\cal F}{\cal F}={\cal F}\tilde {\cal F} = 0$):
	\begin{equation}
		(i(\slashed{\partial} + ie\slashed{\mathcal{A}}) - m)\psi = 0.
	\end{equation}
	Such solutions are called Volkov modes \cite{seipt2017volkov, volkov_1935}, and  our interaction picture Hamiltonian can also be obtained by  simply using (projected)  expansions in Volkov modes. In any case, the interaction-picture interaction Hamiltonian is
	\begin{IEEEeqnarray*}{rCl}
		& & U_0^\dag :\mathrel{H_{\psi A}}: U_0 \supset \int \frac{dp^+ dp^\perp}{(2\pi)^3} \int \frac{dk^+ dk^\perp}{(2\pi)^3\sqrt{k^+}} \sum_{s=\pm\frac{1}{2}} \sum_{j=1,2} em \\
		& \times & \left( \left( \frac{1}{{p'}^+} - \frac{1}{p^+} \right) e^{\frac{ix^+}{2}(k^- + p^- - {p'}^-)} e^{i(f(p)-f(p'))} R^{s,j} c_k^{j\dag} a_p^{s\dag} a_{p'}^{-s} \bigg|_{\mathsf{p}'=\mathsf{p}+\mathsf{k}} \right. \\
		& + & \left( \frac{1}{{p'}^+} - \frac{1}{p^+} \right) e^{\frac{ix^+}{2}(p^- - {p'}^- - k^-)} e^{i(f(p)-f(p'))} R^{s,j} a_p^{s\dag} a_{p'}^{-s} c_k^j \bigg|_{\mathsf{p}'=\mathsf{p}-\mathsf{k}} \\
		& + & \left( \frac{1}{{p'}^+} - \frac{1}{p^+} \right) e^{\frac{ix^+}{2}(k^- + {p'}^- - p^-)} e^{-i(g(p)-g(p'))} R^{-s,j} c_k^{j\dag} b_{p'}^{-s\dag} b_p^s \bigg|_{\mathsf{p}'=\mathsf{p}-\mathsf{k}} \\
		& + & \left( \frac{1}{{p'}^+} - \frac{1}{p^+} \right) e^{\frac{ix^+}{2}({p'}^- - p^- - k^-)} e^{-i(g(p)-g(p'))} R^{-s,j} b_{p'}^{-s\dag} b_p^s c_k^j \bigg|_{\mathsf{p}'=\mathsf{p}+\mathsf{k}} \\
		& + & \left( \frac{1}{{p'}^+} + \frac{1}{p^+} \right) e^{\frac{ix^+}{2}(k^- - p^- - {p'}^-)} e^{-i(g(p)+f(p'))} R^{-s,j} c_k^{j\dag} b_p^s a_{p'}^s \bigg|_{\mathsf{p}'=\mathsf{k}-\mathsf{p}} \\
		& - & \left. \left( \frac{1}{{p'}^+} + \frac{1}{p^+} \right) e^{\frac{ix^+}{2}(p^- + {p'}^- - k^-)} e^{i(f(p)-g(p'))} R^{s,j} a_p^{s\dag} b_{p'}^{s\dag} c_k^j \bigg|_{\mathsf{p}'=\mathsf{k}-\mathsf{p}} \right) \yesnumber
	\end{IEEEeqnarray*}
	Finally, we discretize assuming a momentum lattice with spacing $\Delta p^i = \frac{2\pi }{L}$. Including factors associated with discretizing the ladder operators (see Eq. (\ref{discrete_commutators})), we obtain the final result
	\begin{IEEEeqnarray*}{rCl}
		& & U_0^\dag :\mathrel{H_{\psi A}}: U_0 \supset \sum_{n_p^+,n_p^1,n_p^2} \sum_{n_k^+,n_k^1,n_k^2} \sum_{s=\pm\frac{1}{2}} \sum_{j=1,2} \frac{em}{\sqrt{k^+L^3}} \\
		& \times & \left( \left( \frac{1}{{p'}^+} - \frac{1}{p^+} \right) e^{\frac{ix^+}{2}(k^- + p^- - {p'}^-)} e^{i(f(p)-f(p'))} R^{s,j} c_k^{j\dag} a_p^{s\dag} a_{p'}^{-s} \bigg|_{\mathsf{p}'=\mathsf{p}+\mathsf{k}} \right. \\
		& + & \left( \frac{1}{{p'}^+} - \frac{1}{p^+} \right) e^{\frac{ix^+}{2}(p^- - {p'}^- - k^-)} e^{i(f(p)-f(p'))} R^{s,j} a_p^{s\dag} a_{p'}^{-s} c_k^j \bigg|_{\mathsf{p}'=\mathsf{p}-\mathsf{k}} \\
		& + & \left( \frac{1}{{p'}^+} - \frac{1}{p^+} \right) e^{\frac{ix^+}{2}(k^- + {p'}^- - p^-)} e^{-i(g(p)-g(p'))} R^{-s,j} c_k^{j\dag} b_{p'}^{-s\dag} b_p^s \bigg|_{\mathsf{p}'=\mathsf{p}-\mathsf{k}} \\
		& + & \left( \frac{1}{{p'}^+} - \frac{1}{p^+} \right) e^{\frac{ix^+}{2}({p'}^- - p^- - k^-)} e^{-i(g(p)-g(p'))} R^{-s,j} b_{p'}^{-s\dag} b_p^s c_k^j \bigg|_{\mathsf{p}'=\mathsf{p}+\mathsf{k}} \\
		& + & \left( \frac{1}{{p'}^+} + \frac{1}{p^+} \right) e^{\frac{ix^+}{2}(k^- - p^- - {p'}^-)} e^{-i(g(p)+f(p'))} R^{-s,j} c_k^{j\dag} b_p^s a_{p'}^s \bigg|_{\mathsf{p}'=\mathsf{k}-\mathsf{p}} \\
		& - & \left. \left( \frac{1}{{p'}^+} + \frac{1}{p^+} \right) e^{\frac{ix^+}{2}(p^- + {p'}^- - k^-)} e^{i(f(p)-g(p'))} R^{s,j} a_p^{s\dag} b_{p'}^{s\dag} c_k^j \bigg|_{\mathsf{p}'=\mathsf{k}-\mathsf{p}} \right). \yesnumber
		\label{bw_int_hamiltonian}
	\end{IEEEeqnarray*}
	For  general SFQED processes, the complete mode-expanded Hamiltonian can be derived similarly to the terms discussed in this appendix. For the Breit-Wheeler simulation considered in the main body of the work, only the last two lines of Eq.~(\ref{bw_int_hamiltonian}) were needed. 
	
\end{document}